\documentclass[10pt,journal,letterpaper,compsoc]{IEEEtran}

\usepackage{fancyhdr}

\ifCLASSOPTIONcompsoc
   \usepackage[nocompress]{cite}
\else
   \usepackage{cite}
\fi

\ifCLASSINFOpdf
   \usepackage[pdftex]{graphicx}
\else
   \usepackage[dvips]{graphicx}
\fi

\usepackage[cmex10]{amsmath}

\usepackage{array}

\usepackage{mdwmath}
\usepackage{mdwtab}

\usepackage{eqparbox}

\ifCLASSOPTIONcompsoc
  \usepackage[caption=false,font=footnotesize,labelfont=footnotesize,textfont=footnotesize]{subfig}
\else
  \usepackage[caption=false,font=footnotesize,labelfont=footnotesize,textfont=footnotesize]{subfig}
\fi

\usepackage{caption}
\usepackage{url}

\usepackage{verbatim}
\usepackage{amssymb}
\usepackage{multirow}
\usepackage{bbm}
\usepackage{algorithmic}
\usepackage{algorithm}

\def\twofig{\ifdefined\thesis 0.45\else 0.5\fi}

\newtheorem{lem}{Lemma}
\newtheorem{prop}{Proposition}
\newtheorem{defn}{Definition}

\newcommand{\bb}[1]{\mathbb{#1}}

\newcommand{\A}[1]{\mathcal{#1}}

\newcommand{\fref}[1]{Fig.~\ref{#1}}
\newcommand{\sref}[1]{Section~\ref{#1}}
\newcommand{\tref}[1]{Table~\ref{#1}}
\newcommand{\pref}[1]{Prop.~\ref{#1}}
\newcommand{\lref}[1]{Lemma~\ref{#1}}

\usepackage{color}
\newcommand{\red}[1]{{#1}}

\usepackage[left=.6in,right=.5in,top=.65in,bottom=.5in]{geometry}

\begin{document}

\title{A Metric for DISH Networks:\\Analysis, Implications, and Applications}
\author{Tie Luo,~\IEEEmembership{Member,~IEEE,}
        Vikram Srinivasan,~\IEEEmembership{Member,~IEEE,}
        and~Mehul Motani,~\IEEEmembership{Member,~IEEE}
\IEEEcompsocitemizethanks{\IEEEcompsocthanksitem Tie Luo and Mehul Motani are with the Department of Electrical and Computer Engineering, National University of Singapore, Singapore 117582.
\IEEEcompsocthanksitem Vikram Srinivasan is with Bell Labs Research, Bangalore 560095, India.}}

\IEEEcompsoctitleabstractindextext{%
\begin{abstract}
In wireless networks, node cooperation has been exploited as a data relaying mechanism for decades. However, the wireless channel allows for much richer interaction among nodes. {\red In particular, Distributed Information SHaring (DISH) represents a new improvement to multi-channel MAC protocol design by using a cooperative element at the control plane. In this approach, nodes exchange control information to make up for other nodes' insufficient knowledge about the environment, and thereby aid in their decision making.} To date, what is lacking is a theoretical understanding of DISH. In this paper, we view cooperation as a network resource and evaluate the availability of cooperation, $p_{co}$. We first analyze $p_{co}$ in the context of a multi-channel multi-hop wireless network, and then perform simulations which show that the analysis accurately characterizes $p_{co}$ as a function of underlying network parameters. Next, we investigate the correlation between $p_{co}$ and network metrics such as collision rate, packet delay, and throughput. We find a near-linear relationship between $p_{co}$ and the metrics, which suggests that $p_{co}$ can be used as an appropriate performance indicator itself. Finally, we apply our analysis to solving a channel bandwidth allocation problem, where we derive optimal schemes and provide general guidelines on bandwidth allocation for DISH networks.
\end{abstract}

}

\maketitle

\fancypagestyle{firststyle}
{
   \fancyhf{}
   \fancyhead[L]{IEEE TRANSACTIONS ON MOBILE COMPUTING, vol. 9, no. 3, pp. 376-389, March 2010}
   \fancyfoot[C]{\thepage}
}

\renewcommand{\headrulewidth}{0pt}
\thispagestyle{firststyle}

\IEEEdisplaynotcompsoctitleabstractindextext

\IEEEpeerreviewmaketitle

\section{Introduction} \label{sec:intro-ana}

\IEEEPARstart{C}{ooperative} diversity is not a new concept in wireless communications.  Key ideas and results in cooperative communications can be traced back to the 1970s to van der Meulen \cite{van3term} and Cover \& El Gamal~\cite{cover79}, whose works have spurred numerous studies on this topic from an information-theoretic perspective (e.g., \cite{coopcap06,laneman03,coop03toc1,coop03toc2}) or a protocol-design perspective (e.g., \cite{rdcf05infocom,cmac05globecom,cdmac07icc,comac07jsac}).
To date, cooperation has been intensively studied in various contexts, as a {\em data relaying} mechanism where intermediate nodes help relay packets from a transmitter to a receiver. However, the wireless channel allows for much richer interaction among nodes. {\red Consider another possibility, where nodes exchange control information to make up for other nodes' insufficient knowledge about the environment, and thereby aid in their decision making. This represents a notion of {\em control-plane cooperation}, which augments the conventional understanding of cooperation at the data plane.

This paper specifically studies the distributed flavor of control-plane cooperation, which we call Distributed Information SHaring (DISH). It is a general approach and can be customized into specific mechanisms in different environments. For example in a multi-channel network, due to insufficient knowledge about channel or node status, a node may select a conflicting channel to exchange data, or a transmitter may initiate communication to a receiver who is however on a different channel. In both cases (which are the variants of a {\em multi-channel coordination problem} define later), DISH can be used such that neighboring nodes who possess relevant information share the information to help the particular node avoid the problem. Recently, a DISH-based multi-channel MAC protocol was proposed in \cite{tie06cam,tie09tmc}, and simulations therein demonstrated significant performance gain over other non-cooperative protocols.

However, to date, what is still lacking is a theoretical understanding of DISH. The benefit of DISH is that it can remove the need for multiple transceivers \cite{dca00,nas99,jain01,mup04,mah06} and time synchronization \cite{chen03,mmac04,tmmac07,chma00,chat00,ssch04} in designing multi-channel MAC protocols. This motivates us to understand DISH from a theoretical perspective. In this paper, we provide an analysis of DISH in terms of the {\em availability of cooperation} by viewing cooperation as a network resource. We define a metric $p_{co}$ in the context of a general multi-channel network. It characterizes the availability of cooperation as the probability that a multi-channel coordination problem is identified by any neighboring node (who thus becomes cooperative).
We analyze this metric in multi-hop networks with randomly distributed nodes, and verify the analysis via simulations. It is shown that our analysis accurately characterizes the availability of cooperation as a function of underlying network parameters, and reveals {\em what} factors and {\em how} these factors affect cooperation. This serves as a guide in provisioning a DISH network toward its optimal performance.

We then carry out a detailed investigation of $p_{co}$ with three different contexts of DISH: non-cooperative case, ideal DISH, and real DISH. The investigation shows that $p_{co}$ is a very meaningful performance indicator for DISH networks. Moreover, we apply our analysis to solving a practical channel bandwidth allocation problem, which yields insights different from non-cooperative networks.

Throughput analysis for multi-hop networks is difficult (and still an open problem in general), and it gets even more complicated in a multi-channel context with DISH.  Our approach in this paper is to first look at $p_{co}$ and then correlate it with network performance.  We found a simple relationship between $p_{co}$ and typical performance metrics.

The specific findings of this study are:
\begin{enumerate}
  \item The availability of cooperation is high ($p_{co}>0.7$) in typical cases, which suggests that DISH is feasible to use in multi-channel MAC protocols.
  \item The performance degradation due to an increase in node density can be alleviated due to the simultaneously increased availability of cooperation.
  \item The metric $p_{co}$ increases with packet size for a given {\em bit} arrival rate, but decreases with packet size for a given {\em packet} arrival rate.
  \item Node density and traffic load have opposite effects on $p_{co}$ but node density is the dominating factor. This implies an improved scalability for DISH networks as $p_{co}$ increases with node density.
  \item $p_{co}$ is strongly correlated to network performance and has a near-linear relationship with metrics such as throughput and delay. This may significantly simplify performance analysis for cooperative networks, and suggests that $p_{co}$ be used as an appropriate performance indicator itself.
  \item $p_{co}$ is concave and not monotonic with respect to the ratio between control channel bandwidth and data channel bandwidth; there exists one and only one maximum $p_{co}$ for each given set of parameters.
  \item The optimal bandwidth ratio between the control and a data channel increases with node density but decreases with traffic load. {\red This tells us that, for example, in a sparser network with heavier traffic, the control channel should be allocated less bandwidth for larger $p_{co}$.}
  \item
   In most cases (the number of channels is not too small), {\red to boost the availability of cooperation,} the control channel should be allocated more bandwidth than {\red each} data channel, rather than using {\em smaller} frequency band for a control channel or the usual {\em equal} bandwidth partition as suggested by many other studies.
\end{enumerate}

Part of this work was presented in \cite{tie08mobihoc}. We review related work in \sref{sec:rel}, give the system model in \sref{sec:model-ana} and then the analysis in \sref{sec:analysis}. The analysis is verified in \sref{sec:simu-pco} where we also provide an detailed investigation of $p_{co}$ in different contexts of DISH. Then we present an application of our analysis in \sref{sec:bwalloc}. Finally, \sref{sec:conc} gives concluding remarks.

\section{Related Work}\label{sec:rel}

Multi-channel MAC protocols for ad hoc networks can be categorized into multi-radio schemes and single-radio schemes. In the first category \cite{dca00,nas99,jain01,mup04,mah06}, each node dedicates a radio to monitoring channel usage when the other radio(s) are engaged in communication. In the second category, the mainstream approach is using time synchronization to (i) require nodes to negotiate channels in common time slots \cite{chen03,mmac04,tmmac07} or (ii) hop among channels according to well-known or predictable sequences \cite{chma00,chat00,ssch04}. The first category sacrifices cost-efficiency due to the extra hardware requirement, and the second category underperforms in scalability due to the complexity of time synchronization.

There are three other studies most related to this work. {\red One is CAM-MAC~\cite{tie06cam,tie09tmc} which is the first DISH-based protocol. This multi-channel MAC protocol uses a single transceiver but does not require time synchronization. In the protocol, a transmitter and a receiver perform a handshake on the control channel to reserve a free data channel. If they choose a busy channel or the receiver is not on the control channel (and hence deaf to the transmitter), neighboring nodes who recognize this (via overhearing) may send cooperative messages to inform the transmitter and/or the receiver.
Our later evaluation of real DISH will be based on an adapted version of CAM-MAC.}

The second work, CoopMAC~\cite{comac07jsac}, is also a cooperative MAC protocol but the cooperation refers to data relaying as in many other protocols such as \cite{rdcf05infocom,cmac05globecom,cdmac07icc}. This protocol replaces a single low-rate transmission with two high-rate transmissions by using a relay node, in order to achieve higher throughput. The paper provides a protocol analysis which involves computing the probability that a relay node is available. This probability seems to be similar to $p_{co}$ but they are actually very different.
The probability defined in \cite{comac07jsac} is a {\em spatial} metric because it is completely determined by the (static) node placement, i.e., a relay is deemed available if there is a node in the intersection of the radio footprints of the source and the destination. This can be easily calculated via simple geometric analysis (as the paper assumes uniformly distributed nodes). On the other hand, $p_{co}$ is a metric involving spatial, temporal, and frequency factors: besides node locations, nodes' communication activities (which vary with time), and the channels that nodes are tuned to (while \cite{comac07jsac} assumes a single channel), all will affect $p_{co}$. Furthermore, the problem context of the analysis in \cite{comac07jsac} is a {\em wireless LAN} whereas our context is a multi-hop network.
Finally, \cite{comac07jsac} deals with data-plane cooperation whereas our work deals with control-plane cooperation.

The last work is by Han et al. \cite{han05} who considered a multi-channel MAC protocol that adopts ALOHA on the control channel to reserve data channels.  Queueing theory was applied to calculate the throughput of the protocol. However, there are some noteworthy limitations. First, the scenario under consideration is single-hop. Second, it is assumed that each node can communicate on the control channel and a data channel simultaneously. This implicitly requires {\em two transceivers} per node and consequently leads to collision-free data channels. Third, a unique {\em virtual queue} was assumed to store the packets arriving at {\em all} nodes for the purpose of {\em centralized transmission scheduling}, and the precise status of this queue was assumed to be known to the entire network. This assumption is impractical and makes the calculated throughput actually an upper bound. Fourth, the access to the control channel uses the ALOHA mechanism while CSMA/CA is widely adopted in practice.

\section{System Model}\label{sec:model-ana}

We consider a static and connected ad hoc network in which each node is equipped with a single half-duplex transceiver that can dynamically switch between a set of orthogonal frequency channels but can only use one at a time. One channel is designated as the control channel and the others as data channels. Nodes are placed in a plane area according to a two-dimensional Poisson point process.

We consider a class of multi-channel MAC protocols with their common framework described below. A transmitter-receiver pair uses an \texttt{McRTS}/\texttt{McCTS} handshake on the control channel to set up communication (like 802.11 RTS/CTS) for their subsequent \texttt{DATA}/\texttt{ACK} handshake on a data channel. To elaborate, a transmitter sends an \texttt{McRTS} on the control channel using CSMA/CA, i.e., it sends \texttt{McRTS} after sensing the control channel to be idle for a random period (addressed below) of time. The intended receiver, after successfully receiving \texttt{McRTS}, will send an \texttt{McCTS} and then switch to a data channel (the \texttt{McRTS} informs the receiver of the chosen data channel).
After successfully receiving the \texttt{McCTS}, the transmitter will also switch to its chosen data channel, and otherwise it will back off on the control channel for a random period (addressed below) of time. Hence it is possible that only the receiver switches to the data channel. After switching to a data channel, the transmitter will send a \texttt{DATA} and the receiver will respond with a \texttt{ACK} upon successful reception. Then both of them switch back to the control channel.

In the above we have mentioned two random periods. They are assumed to be designed such that idle intervals on the control channel are well randomized. Specifically, when a node is on the control channel, it sends control messages (an aggregated stream of \texttt{McRTS} and \texttt{McCTS}) according to a Poisson process with an unknown average rate (will be determined in \sref{sec:analysis}). The validity of this assumption will be verified via simulations.

Note that we use \texttt{McRTS}, \texttt{McCTS}, \texttt{DATA} and \texttt{ACK} to refer to different packets (frames) without assuming specific frame formats. Since, logically, they must make a protocol functional, we assume that \texttt{McRTS} carries channel usage information (e.g., ``who will use which channel for how long'') and, for simplicity, \texttt{McCTS} is the same as \texttt{McRTS}.

We assume that, after switching to a data channel, a node will stay on that channel for a period of $T_d$, where $T_d$ is the duration of a successful data channel handshake.\ifdefined\thesis\ \input{../quantify/pco-td}

\else\footnote{This is also valid for a failed data channel handshake. See appendix for the elaboration.}\ \fi
We ignore channel switching delay as it will not fundamentally change our results if it is negligible compared to $T_d$ (the delay is $80\mu s$\cite{ssch04} while $T_d$ is more than $6ms$ for a 1.5KB data packet on a 2Mb/s channel). We also ignore SIFS and propagation delay for the same reason, provided that they are smaller than the transmission time of a control message.

We assume a uniform traffic pattern --- all nodes have the equal data packet arrival rate, and for each data packet to send, a node chooses a receiver equally likely among its neighbors.
We also assume a stable network --- all data packets can be delivered to destinations within finite delay. In addition, packet reception fails if and only if packets collide with each other (i.e., no capture effect), transmission and interference ranges are equal, and neighboring nodes do not start sending control messages simultaneously (there is no time synchronization).

We do not assume a specific channel selection strategy; how a node selects data channels will affect how often conflicting channels are selected, but will not affect $p_{co}$. This is because, intuitively, we only care about the availability of cooperation ($p_{co}$) when a multi-channel coordination problem (a precise definition is given in Def.~\ref{def:mccp-ana}), which includes channel conflicting problem, {\em has been} created.

We do not assume a concrete DISH mechanism, i.e., nodes do not physically react upon a multi-channel coordination problem, because analyzing the availability of cooperation does not require the use of this resource. In fact, assuming one of the (numerous possible) DISH mechanisms will lose generality. Nevertheless, we will show in \sref{sec:simu-pco} that, when an ideal or a real DISH mechanism is used, the results do not fundamentally change. This could be an overall effect from contradicting factors which will be explained therein.

We assume that the following parameters are known:
\begin{itemize}
\item $n$: node density. In a multi-hop network, it is the average number of nodes per $R^2$ where $R$ is the transmission range. In a single-hop network, it is the total number of nodes.
\item $\lambda$: the average data packet arrival rate at each node, including retransmissions.
\item $T_d$: the duration of a data channel handshake.
\item $b$: the transmission time of a control message. $b\ll T_d$.
\end{itemize}

\renewcommand{\labelenumi}{(\alph{enumi})}

\section{Analysis} \label{sec:analysis}

\subsection{Problem Formulation and Analysis Outline}\label{sec:formulate}

We first formally define $p_{co}$, which depends on two concepts called the {\em MCC problem} and the {\em cooperative node}.

\begin{defn}[MCC Problem]\label{def:mccp-ana}
A multi-channel coordination (MCC) problem is either a {\em channel conflict} problem or a {\em deaf terminal} problem. A channel conflict problem is created when a node, say $y$, selects a channel to use (transmit or receive packets) but the channel is already in use by a neighboring node, say $x$. A deaf terminal problem is created when a node, say $y$, initiates communication to another node, say $x$, that is however on a different channel. In either case, we say that an MCC problem is created by $x$ and $y$.
\end{defn}

In a protocol that transmits {\tt DATA} without requiring {\tt ACK}, a channel conflict problem does not necessarily indicate an impending data collision. We do not consider such a protocol.

\begin{figure}[tb]
\centering
\includegraphics[width=0.8\linewidth]{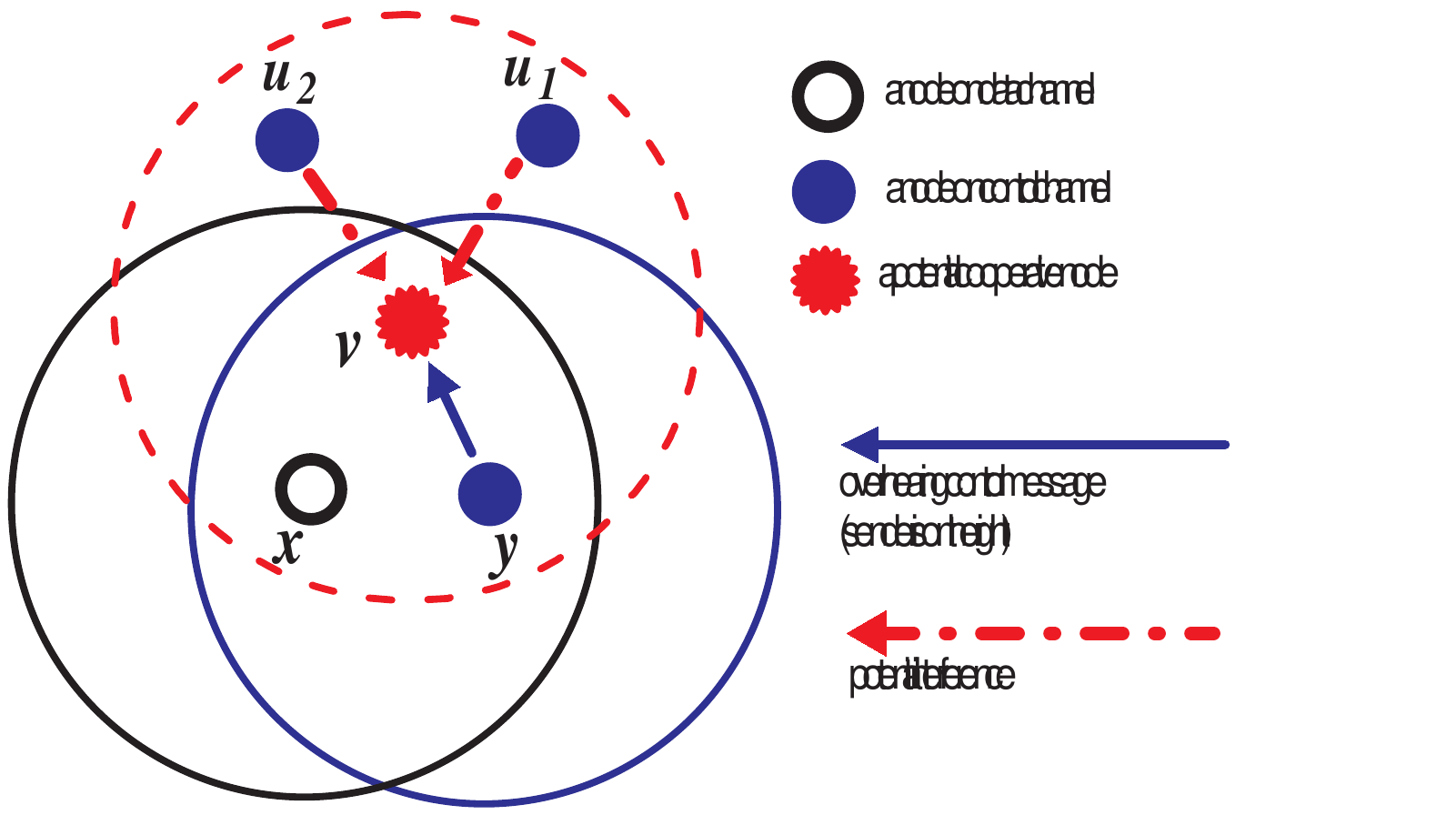}
\caption{Illustration of an MCC problem and a cooperative node. Node $x$ is performing a data channel handshake on $CH_x$, and $y$ has just sent a control message during a control channel handshake. If this control message is an {\tt McRTS} addressed to $x$, then a {\em deaf terminal} problem is created. If this control message indicates that $y$ selects $CH_x$ (recall that a control message carries channel usage information), a {\em channel conflict} problem is created. In either case, if a third node $v$ identifies this problem (by overhearing $x$'s and $y$'s control messages successively), it is a cooperative node.}
\label{fig:coopnode}
\vspace{-4mm}\end{figure}

\begin{defn}[Cooperative Node]\label{def:coopnode}
A node that identifies an MCC problem created by two other nodes, say $x$ and $y$, is called a cooperative node with respect to $x$ and $y$.
\end{defn}

\fref{fig:coopnode} gives a visualization of the above two concepts in our system model.

\begin{table*}[tb]
\caption{Notation}\label{tab:notation}
\centering
\ifdefined\thesis
\begin{tabular} {| c | l | p{4.5in} | }
\else
\begin{tabular} {| c | l | p{5.6in} | }
\fi
\hline
\multirow{5}{*}{\rotatebox{90}{\mbox{Probabilities}}}
& $p_{co}^{xy}$ & the probability that at least one cooperative node with respect to $x$ and $y$ exists \\
& $p_{co}^{xy}(v)$ & the probability that node $v$ is a cooperative node with respect to $x$ and $y$\\
& $p_{ctrl}$ & the probability that a node is on the control channel at an arbitrary point in time \\
& $p_{succ}$ & the probability that a control channel handshake (initiated by an \texttt{McRTS}) is successful \\
& $p_{oh}$ & the probability that an arbitrary node successfully overhears a control message \\
\hline
\multirow{6}{*}{\rotatebox{90}{\mbox{Events}}}
& $\A C_v(t)$ & node $v$ is on the control channel at time $t$ \\
& $\A O (v\leftarrow i)$ & node $v$ successfully overhears node $i$'s control message, given that $i$ sends the message \\
& $\A S_v(t_1,t_2)$ & node $v$ is silent (not transmitting) on the control channel during interval $[t_1, t_2]$ \\
& $\A I_v(t_1,t_2)$ & node $v$ does not introduce interference to the control channel during interval $[t_1, t_2]$, i.e., it is on a data channel or is silent on the control channel\\
& $\Omega_u(t_1,t_2)$ & node $u$, which is on a data channel at $t_1$, switches to the control channel in $[t_1,t_2]$ \\
\hline
\multirow{5}{*}{\rotatebox{90}{\mbox{Others}}}
& $\A N_i, \A N_{ij}, \A N_{v\backslash i}$ & $\A N_i$ is the set of node $i$'s neighbors,
$\A N_{ij}=\A N_i \cap \A N_j$, $\A N_{v\backslash i}=\A N_v\backslash \A N_i\backslash \{i\}$ ($v$'s but not $i$'s neighbors) \\
& $K_{ij}, K_{v\backslash i}$ & $K_{ij}=|\A N_{ij}|$, $K_{v\backslash i}=|\A N_{v\backslash i}|$ \\
& $s_i$ & the time when node $i$ starts to send a control message\\
& $\lambda_c, \lambda_{rts}, \lambda_{cts}$ & the average rates of a node sending control messages, \texttt{McRTS}, and \texttt{McCTS}, respectively, {\em when it is on the control channel}. Clearly, $\lambda_c = \lambda_{rts}+\lambda_{cts}$\\
\hline
\end{tabular}
\end{table*}

\begin{defn}[$p_{co}$]\label{def:pco}
$p_{co}$ is the probability for two arbitrary nodes that create an MCC problem to obtain cooperation, i.e., there is at least one cooperative node with respect to these two nodes.
\end{defn}

Note that if there are multiple cooperative nodes and they are allowed by a DISH mechanism to send cooperative messages concurrently, a collision can result at node $y$. However, this collision {\em still indicates an MCC problem} and thus cooperation is still obtained. CAM-MAC\cite{tie06cam,tie09tmc} also implements this.

We distinguish the receiving of control messages. A transmitter receiving \texttt{McCTS} from its intended receiver is referred to as {\em intentional receiving}, and the other cases of receiving are referred to as {\em overhearing}, i.e., any node receiving \texttt{McRTS} (hence an intended receiver may also be a cooperative node) or any node other than the intended transmitter receiving \texttt{McCTS}.

Our notation is listed in \tref{tab:notation}. Overall, we will determine $p_{co}$ by following the order of $p_{co}^{xy}(v)\rightarrow p_{co}^{xy}\rightarrow p_{co}$.

Consider $p_{co}^{xy}(v)$ first. \fref{fig:coopnode} illustrates that node $v$ is cooperative if and only if it successfully overhears $x$'s and $y$'s control messages successively. Hence $\forall v\in \A N_{xy}$,
\begin{align}\label{eq:pcoxyv-def}
p_{co}^{xy}(v) &= \Pr[\A O(v\leftarrow x),\A O(v\leftarrow y)]\notag\\
&= \Pr[\A O (v\leftarrow x)]\cdot \Pr[\A O (v\leftarrow y)|\A O (v\leftarrow x)].
\end{align}

Consider $\A O (v\leftarrow i)$. For $v$ to successfully overhear $i$'s control message which is being sent during interval $[s_i,s_i+b]$, $v$ must be silent on the control channel and not be interfered, i.e.,
\begin{align}\label{eq:poh-def}
\Pr[\A O (v\leftarrow i)] =& \Pr[\A S_v(s_i, s_i+b),
    \bigcap_{u \in \A N_v\backslash \{i\}} \A I_u(s_i,s_i+b)],\notag\\
    &\forall v \in \A N_i.
\end{align}

Now we outline our analysis as below.
\begin{itemize}
\item \sref{sec:solve-poh}: {\red solves \eqref{eq:poh-def}.}
\item \sref{sec:revisit}: solves \eqref{eq:pcoxyv-def} and the target metric $p_{co}$.
\item \sref{sec:sghop}: case study in single-hop networks.
\end{itemize}
\ifdefined\thesis
\else
{\red Proofs are in the Appendix unless otherwise specified.}
\fi

\subsection{{\red Solving \eqref{eq:poh-def}}} \label{sec:solve-poh}

\ifdefined\thesis
\input{../quantify/pco-lem-switch}
\fi

\begin{prop}\label{prop:notintfoh}
If node $v$ is overhearing a control message from node $i$ during $[s_i, s_i+b]$, then the probability that a node $u \in \A N_v$ does not interfere with $v$ is given by
\begin{align*}
\Pr[\A I_u(s_i, s_i+b)] = \left\{
  \begin{array}{ll}
    1, & u \in \A N_{vi}; \\
    p_{ni\text{-}oh}, & u \in \A N_{v\backslash i}.
  \end{array}
\right.
\end{align*}
where
\[ p_{ni\text{-}oh} = \ p_{ctrl} \cdot e^{-2 \lambda_c b} +(1-p_{ctrl})
    \cdot (1-\frac{2b}{T_d} + \frac{1- e^{-2 \lambda_c b}}{\lambda_c T_d}). \]
\end{prop}

\ifdefined\thesis
\input{../quantify/pco-proof-nioh}
\fi

Thus \eqref{eq:poh-def} can be reduced to \ifdefined\thesis\else(see Appendix)\fi
\begin{align}\label{eq:poh-vi}
\Pr[\A O (v\leftarrow i)] \approx p_{ctrl} \; p_{ni\text{-}oh}^{K_{v\backslash i}}.
\end{align}

\ifdefined\thesis
\input{../quantify/pco-proof-eq1}
\fi

{\red The above contains two unknown variables, $p_{ctrl}$ and $\lambda_c$, and the following solves for them.}

For $p_{ctrl}$, consider a sufficiently long period $T_0$. On the one hand, the number of arrival data packets at each node is $\lambda T_0$. On the other hand, each node spends a total time of $(1-p_{ctrl})T_0$ on data channels, a factor $\eta$ of which is used for sending arrival data packets. Since the network is stable (incoming traffic is equal to outgoing traffic), we establish a balanced equation:
\begin{align*}
    \lambda T_0 T_d = \eta\; (1-p_{ctrl}) T_0.
\end{align*}
To determine $\eta$, noticing that a node switches to data channels either as a transmitter (with an average rate of $\lambda$) or as a receiver (with an average rate of $\lambda_{cts}$), we have
$\eta =\lambda/(\lambda+\lambda_{cts})$. Substituting this into the above yields
\begin{align}\label{eq:pctrl}
p_{ctrl} = 1 - (\lambda+\lambda_{cts}) T_d.
\end{align}

{\red For $\lambda_c$ (and $\lambda_{cts}$), we need a proposition and two lemmas.}
\begin{prop}\label{prop:notintfcts}
If node $i$ (transmitter) is intentionally receiving \texttt{McCTS} from node $j$ (receiver) during $[s_j, s_j+b]$, then the probability that a node $u \in \A N_i$ does not interfere with $i$ is given by
\begin{align*}
\Pr[\A I_u(s_j, s_j+b)] = \left\{
  \begin{array}{ll}
    1, & u \in \A N_{ij}; \\
    p_{ni\text{-}cts}, & u \in \A N_{i\backslash j}.
  \end{array}
\right.
\end{align*}
where $p_{ni\text{-}cts} =$
\[ (1-p_{ctrl}) [1-\frac{b}{T_d}
       (1+\frac{b}{T_d} - \frac{1- e^{-\lambda_c b}}{\lambda_c T_d} - e^{-\lambda_c b})]
    + p_{ctrl}. \]
\end{prop}

\ifdefined\thesis
\input{../quantify/pco-proof-nicts}
\fi
\begin{lem} \label{lem:epk}
For a Poisson random variable $K$ with mean value $\overline{K}$, and $0<p<1$,
\begin{equation*}
    \bb{E} [p^K] = e^{-(1-p)\overline{K}}.
\end{equation*}
\end{lem}

\begin{IEEEproof}
\begin{align*}
    \bb{E} [p^K] = \sum_0^\infty p^k \Pr(K=k)
        = e^{-\overline{K}} \sum_0^\infty \frac{(p\overline{K})^k}{k!}
        = e^{-(1-p)\overline{K}}.
\end{align*}
\end{IEEEproof}

\begin{lem} \label{lem:avgarea}
For three random distributed nodes $v$, $i$ and $j$,
\begin{enumerate}
\item $\bb{E}[K_{v\backslash i}|v\in \A N_i] \approx 1.30 n$.
\item $\bb{E}[K_{v\backslash i}|v\in \A N_{ij}] \approx 1.19 n$.
\item $\bb{E}[K_{ij}] \approx 1.84 n$.
\end{enumerate}
\end{lem}

\ifdefined\thesis
\input{../quantify/pco-proof-avgarea}
\fi

Now we can prove \ifdefined\thesis\else(see Appendix)\fi
\begin{align}\label{eq:poh_psucc}
p_{oh} \approx p_{ctrl}\; \exp[-1.30 n (1-p_{ni\text{-}oh})],\notag\\
p_{succ} \approx p_{oh} \exp[-1.30 n(1-p_{ni\text{-}cts})].
\end{align}
\ifdefined\thesis
\input{../quantify/pco-proof-eq2}
\fi

Now we solve $\lambda_c$ (together with $\lambda_{cts}$). From the perspective of a transmitter, the average number of successful control channel handshakes that it initiates per second is $p_{ctrl} \lambda_{rts} p_{succ}$. Since each successful control channel handshake leads to transmitting one data packet, we have
\[ p_{ctrl} \lambda_{rts} p_{succ} = \lambda.\]

From the perspective of a receiver, it sends an \texttt{McCTS} when it successfully receives (overhears) an \texttt{McRTS} addressed to it, \footnote{In a more sophisticated protocol model, a receiver may not respond even if it receives McRTS due to, e.g., disagreeing with the transmitter's channel selection.
This behavior is modeled by ideal DISH and real DISH, which will be shown in \sref{sec:simu-pco} that it does not fundamentally change the results.}
and hence $\lambda_{cts} = \lambda_{rts}\, p_{oh}$.  Then combining these with $\lambda_c = \lambda_{rts}+\lambda_{cts}$ yields
\begin{align}\label{eq:lmdc}
    \lambda_c = \frac{\lambda (1+p_{oh})}{p_{ctrl}\, p_{succ}} \text{ and }
    \lambda_{cts} = \frac{\lambda\ p_{oh}}{p_{ctrl}\, p_{succ}}
\end{align}
where $p_{oh}$ and $p_{succ}$ are given in \eqref{eq:poh_psucc}.

\subsection{Solving \eqref{eq:pcoxyv-def} and Target Metric $p_{co}$}\label{sec:revisit}

Based on the proof of \eqref{eq:poh-vi}, it can be derived that
\begin{align}\label{eq:poh-vy}
\Pr[\A O (v\leftarrow y)|\A O (v\leftarrow x)] \approx p_{ctrl}^{\star} \; p_{ni\text{-}oh}^{K_{v\backslash y}},
\end{align}
\[ \text{where \ \ } p_{ctrl}^{\star} \triangleq \Pr[\A C_v(s_y)|\A O (v\leftarrow x)]. \]
Note that $p_{ctrl}^{\star} \neq p_{ctrl}$, because $s_y$ is not an {\em arbitrary} time for $v$ due to the effect form $\A O (v\leftarrow x)$. The reason is that $\A O (v\leftarrow x)$ implies $\A C_v(s_x)$, and thus for $\A C_v(s_y)$ to happen, $v$ must stay {\em continuously} on the control channel during $[s_x,s_y]$ (otherwise, a switching will lead to $v$ staying on the data channel for $T_d$, but $s_x+T_d>s_y$ since $x$'s data communication is still ongoing at $s_y$, and hence $\A C_v(s_y)$ can never happen).

It can be proven that \ifdefined\thesis\else(see Appendix)\fi
\begin{align}\label{eq:pctrlstar}
p_{ctrl}^{\star} = \frac{(w\lambda_c - \frac{1-w}{T_d})\; g(\lambda_c+\lambda_w) +
    \frac{1-w}{T_d}\; g(\lambda_w)} {1-w+ (w\lambda_c - \frac{1-w}{T_d})\; g(\lambda_c)}
\end{align}
where
\[ g(x)=\frac{1- e^{-x T_d}}{x},\;\; w = \frac{p_{ctrl}-p_{oh}}{1-p_{oh}}, \text{ and}\]
\[ \lambda_w=\lambda_{rts} p_{succ}+\lambda_{cts}. \]

\ifdefined\thesis
\input{../quantify/pco-proof-eq3}
\fi

Combining \eqref{eq:poh-vi} and \eqref{eq:poh-vy} reduces \eqref{eq:pcoxyv-def} to
\begin{align}\label{eq:pcoxyv}
p_{co}^{xy}(v) \approx p_{ctrl} \; p_{ctrl}^{\star}
  \; p_{ni\text{-}oh}^{K_{v\backslash x}+K_{v\backslash y}},\; \forall  v\in\A N_{xy}.
\end{align}

Let $p_{co}^{xy}(\star)$ be the average of $p_{co}^{xy}(v)$ over all $v\in\A N_{xy}$, i.e., $p_{co}^{xy}(\star)$ is the probability that an arbitrary node in $\A N_{xy}$ is cooperative with respect to $x$ and $y$, Using \lref{lem:avgarea}-(b),
\begin{align}\label{eq:pcostar}
p_{co}^{xy}(\star) \approx p_{ctrl} \; p_{ctrl}^{\star} \; \exp[-2.38 n (1-p_{ni\text{-}oh})].
\end{align}

By the definition of $p_{co}^{xy}$ in \tref{tab:notation},
\begin{align}\label{eq:pcoxy-approx}
    p_{co}^{xy}
    \approx 1- \prod_{v\in \A N_{xy}} [ 1- p_{co}^{xy}(v) ]
    \approx 1- [ 1- p_{co}^{xy}(\star) ]^{K_{xy}},
\end{align}
where the events corresponding to $1- p_{co}^{xy}(v)$, i.e., nodes not being cooperative with respect to $x$ and $y$, are regarded as independent of each other, as an approximation.

Thus $p_{co}$ is determined by averaging $p_{co}^{xy}$ over all $(x,y)$ pairs that are possible to create MCC problems.
It can be proved \cite{tie07mobicom} that these pairs are neighboring pairs $(x,y)$ satisfying ($d_i$ denoting the degree of a node $i$)
\begin{enumerate}
\item $d_x\ge 2$, $d_y\ge 2$, but not $d_x=d_y=2$, or
\item $d_x=d_y=2$, but $x$ and $y$ are not on the same three-cycle (triangle).
\end{enumerate}
This condition is satisfied by all neighboring pairs in a {\em connected} random network, because the connectivity requires a sufficiently high node degree ($5.18\log N$ where $N$ is the total number of nodes\cite{xue04conn}) which is much larger than 2. Therefore, taking expectation of \eqref{eq:pcoxy-approx} over all neighboring pairs using \lref{lem:epk} and \lref{lem:avgarea}-(c),
\begin{align}\label{eq:pcoxy}
p_{co} &= 1- \exp[ -p_{co}^{xy}(\star) \overline{K_{xy}}]\notag\\
  &\approx 1- \exp[ -1.84 n\; p_{co}^{xy}(\star) ].
\end{align}
This completes the analysis.

\subsection{Special Case: Single-Hop Networks}\label{sec:sghop}

Now that all nodes are in the communication range of each other, we have $p_{ni\text{-}oh}=p_{ni\text{-}cts}=1$ according to \pref{prop:notintfoh} and \ref{prop:notintfcts}, which leads to $p_{succ}=p_{oh}=p_{ctrl}$ according to \eqref{eq:poh_psucc}, and $p_{co}^{xy}(v)= p_{ctrl} \; p_{ctrl}^{\star}$ according to \eqref{eq:pcoxyv}. Hence \eqref{eq:pcoxy-approx} reduces to
\begin{align*}
p_{co}^{xy} = 1- (1- p_{ctrl} \; p_{ctrl}^{\star})^{K_{xy}},
\end{align*}
where $K_{xy}$ is the number of all possible cooperative nodes with respect to $x$ and $y$, leading to $K_{xy}=n-4$.~\footnote{This means all nodes excluding $x$, $x'$, $y$ and $y'$, where $x'$ is the node that $x$ is currently communicating with (on a data channel), and $y'$ is the node that $y$ {\em was} communicating with (on a data channel) when $x$ was setting up communication with $x'$ (hence $y$ and $y'$ missed $x$'s control message). None of these four nodes can be a cooperative node.}
So, as the average of $p_{co}^{xy}$,
\begin{align}\label{eq:sghop}
    p_{co} = 1- (1- p_{ctrl}\; p_{ctrl}^{\star})^{n-4},
\end{align}
where $p_{ctrl}$ is given below, by solving equations \eqref{eq:pctrl}, \eqref{eq:poh_psucc} and \eqref{eq:lmdc},
\begin{align*}
p_{ctrl} &= \frac{1}{2} (1 - \lambda T_d +
    \sqrt{1+\lambda T_d (\lambda T_d - 6)}\ ),
\end{align*}
and $p_{ctrl}^{\star}$ is given below, by reducing \eqref{eq:pctrlstar} with $w=0$,
\[ p_{ctrl}^{\star} = \frac{g(\lambda_w) - g(\lambda_c+\lambda_w)} {T_d - g(\lambda_c)} \]
\begin{align*}
\text{where \ }
\lambda_c &= \frac{1}{2} ( \frac{1- \sqrt{1+\lambda T_d (\lambda T_d - 6)}}
    {\lambda T_d^2} -\frac{3}{T_d}),\\
\lambda_w &= \frac{1- \sqrt{1+\lambda T_d (\lambda T_d - 6)}}{T_d} -\lambda.
\end{align*}

\section{Investigating $p_{co}$ with DISH}\label{sec:simu-pco}

We verify the analysis in both single-hop and multi-hop networks and identify key findings therein. We also investigate the correlation between $p_{co}$ and network performance.

\subsection{Protocol Design and Simulation Setup}\label{sec:proto-pco}

\subsubsection{{\red Non-Cooperative Case}}
This is a multi-channel MAC protocol based on the protocol framework described in \sref{sec:model-ana}. Key part of its pseudo-code is listed below, where $S_{ctrl}$ is the control channel status (FREE/BUSY) detected by the node running the protocol, $S_{node}$ is the node's state (IDLE/TX/RX, etc.), $L_{queue}$ is the node's current queue length, and they are initialized as FREE, IDLE and 0, respectively.
The frame format of \texttt{McRTS} and \texttt{McCTS} is shown in \fref{fig:format-pco}, where we can see that they carry channel usage information. A node that overhears \texttt{McRTS} or \texttt{McCTS} will cache the information in a {\em channel usage table} shown in \fref{fig:chtab}, where {\tt Until} is converted from \texttt{Duration} by adding the node's own clock.

\ifdefined\thesis
\begin{figure}[tb]
\begin{minipage}[b]{0.6\linewidth}
\includegraphics[trim=1mm 2mm 2mm 1mm,clip,width=\linewidth]{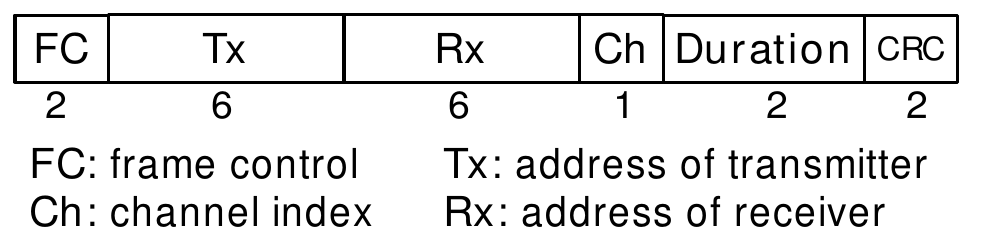}
\caption{Frame format of {\tt McRTS} and {\tt McCTS}.}\label{fig:format-pco}
\end{minipage}\hfil
\begin{minipage}[b]{0.33\linewidth}
\rightline{
\includegraphics[width=\linewidth]{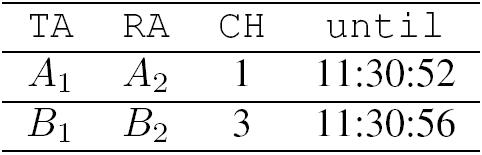}}
\caption{Channel usage table.}\label{fig:chtab-pco}
\end{minipage}
\end{figure}
\else
\begin{figure}[tb]
\center
\includegraphics[trim=1mm 2mm 2mm 1mm,clip,width=0.7\linewidth]{format_rtscts_new}
\caption{Frame format of {\tt McRTS} and {\tt McCTS}.}\label{fig:format-pco}
\end{figure}
\begin{figure}[tb]
\center
\includegraphics[width=0.4\linewidth]{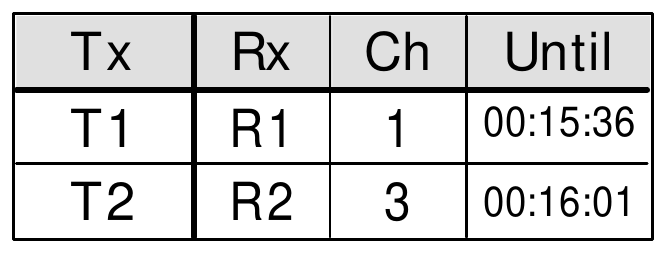}
\caption{Channel usage table.}\label{fig:chtab}
\end{figure}
\fi

\floatname{algorithm}{Procedure}

\begin{algorithm}[htb]
\caption{PKT-ARRIVAL}[Called when a data packet arrives]
\label{alg:pktarr}
\begin{algorithmic}[1]
\STATE enqueue the packet, $L_{queue}$++
\IF{$S_{ctrl}=FREE \wedge S_{node}=IDLE \wedge L_{queue}=1$}
\STATE call ATTEMPT-RTS
\ENDIF
\end{algorithmic}
\end{algorithm}

\begin{algorithm}[htb]
\caption{ATTEMPT-RTS}[Called by PKT-ARRIVAL or CHECK-QUEUE]
\label{alg:rts}
\begin{algorithmic}[1]
\STATE construct a set $\A F$ of free channel indexes using channel usage table
\IF{$\A F\neq\phi$}
\STATE send \texttt{McRTS} with \texttt{CH}$:=$RANDOM$(\A F)$
\ELSE
\STATE Timer $\leftarrow \min(\texttt{until}-now)$
\WHILE{$S_{ctrl}=FREE\ \wedge$ Timer not expired}
\STATE wait \COMMENT{carrier sensing remains on}
\ENDWHILE
\IF{Timer expired}
\STATE call CHECK-QUEUE
\ELSE
\STATE call PASSIVE \COMMENT{receive a control message}
\ENDIF
\ENDIF
\end{algorithmic}
\end{algorithm}

\begin{algorithm}[htb]
\caption{CHECK-QUEUE}[Called when $S_{ctrl}=FREE \wedge S_{node}=IDLE$ changes from $FALSE$ to $TRUE$]
\label{alg:sendca}
\begin{algorithmic}[1]
\IF{$L_{queue}>0$}
\STATE Timer $\leftarrow$ RANDOM$(0,10b)$ \COMMENT{FAMA\cite{fama95,fama99}}
\WHILE{$S_{ctrl}=FREE\ \wedge$ Timer not expired}
\STATE wait \COMMENT{carrier sensing remains on}
\ENDWHILE
\IF{Timer expired}
\STATE call ATTEMPT-RTS
\ELSE
\STATE call PASSIVE \COMMENT{receive a control message}
\ENDIF
\ENDIF
\end{algorithmic}
\end{algorithm}

\ifdefined\EA
\begin{wrapfigure}{r}{0.4\textwidth}
  \begin{center}
    \includegraphics[trim=1mm 2mm 1mm 2mm,clip,width=0.35\textwidth]{fig/format_rtscts}
  \end{center}
  \caption{The frame format of \texttt{McRTS} and \texttt{McCTS}.}
  \label{fig:format}
\end{wrapfigure}
A node that overhears \texttt{McRTS} or \texttt{McCTS} (the frame format is shown in \fref{fig:format}) will cache the channel usage information carried by the packet in a {\em channel usage table}. Each entry of this table has four fields: \texttt{TA}, \texttt{RA}, \texttt{CH}, and \texttt{until}, where \texttt{until} is converted from \texttt{duration} (in the packet) by adding to the node's current time.
\fi

As is based on the system model, this protocol does not use a concrete DISH mechanism, i.e., cooperation is treated as a resource while not actually utilized.

\subsubsection{Ideal DISH}

This protocol is by adding an ideal cooperating mechanism to the {\red non-cooperative case}. Each time when an MCC problem is created by nodes $x$ and $y$ and if at least one cooperative node is available, the node that is on the control channel, i.e., node $y$, will be informed without any message actually sent, and then back off to avoid the MCC problem.

\subsubsection{Real DISH}

{\red In this protocol, cooperative nodes actually send cooperative messages to inform the transmitter or receiver of an MCC problem so that it will back off. We design a real DISH protocol by adapting CAM-MAC\cite{tie06cam}
\ifdefined \thesis
. %
\else
and show the control channel handshake in \fref{fig:hsk-cam}. A transmitter and a receiver first perform a PRA/PRB exchange like the 802.11 RTS/CTS to negotiate a data channel, and then perform a CFA/CFB exchange to confirm (to their respective neighbors) the channel selection. A cooperative node sends an INV packet when identifying an MCC problem through PRA or PRB.
\begin{figure}[tb]
\centering
\includegraphics[trim=3.7cm 12cm 3.7cm 12cm,clip,width=0.8\linewidth]{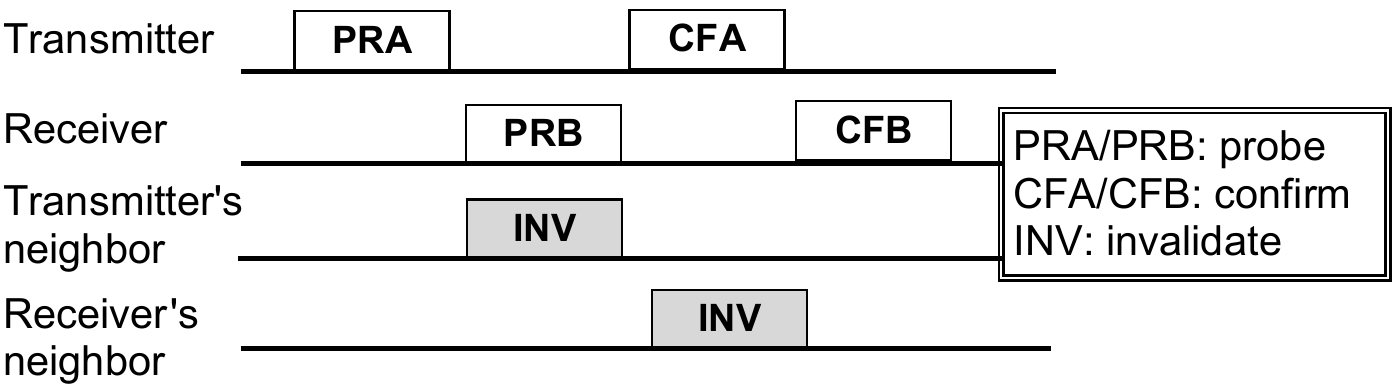}
\caption{{\red Control channel handshake of Real DISH.}}
\label{fig:hsk-cam}
\vspace{-4mm}\end{figure}
\fi
We change CAM-MAC such that $\|PRA\|+\|CFA\|=\|McRTS\|=\|PRB\|+\|CFB\|=\|McCTS\|$, where $\|\cdot\|$ denotes packet size.}

\subsubsection{Simulation Setup}

There are six channels of data rate 1Mb/s each (the number of channels does not affect results as long as the network is kept stable).
Data packets arrive at each node as a Poisson process.
The uniform traffic pattern as in the model is used.
Traffic load $\lambda$ (pkt/s), node density $n$ (1/$R^2$), and packet size $L$ (byte) will vary in simulations. In multi-hop networks, the network area is 1500m$\times$1500m and the transmission range is 250m. Each simulation is terminated when a total of 100,000 data packets are sent over the network, and each set of results is averaged over 15 randomly generated networks.

\subsection{Investigation with {\red Non-Cooperative Case}}\label{sec:modeldish}

\begin{figure}[tb]
\centerline{
\subfloat[$p_{co}$ versus $\lambda$ ($n=8$), single-hop.]
    {\includegraphics[trim=4mm 1mm 11mm 7mm,clip,width=\twofig\linewidth]{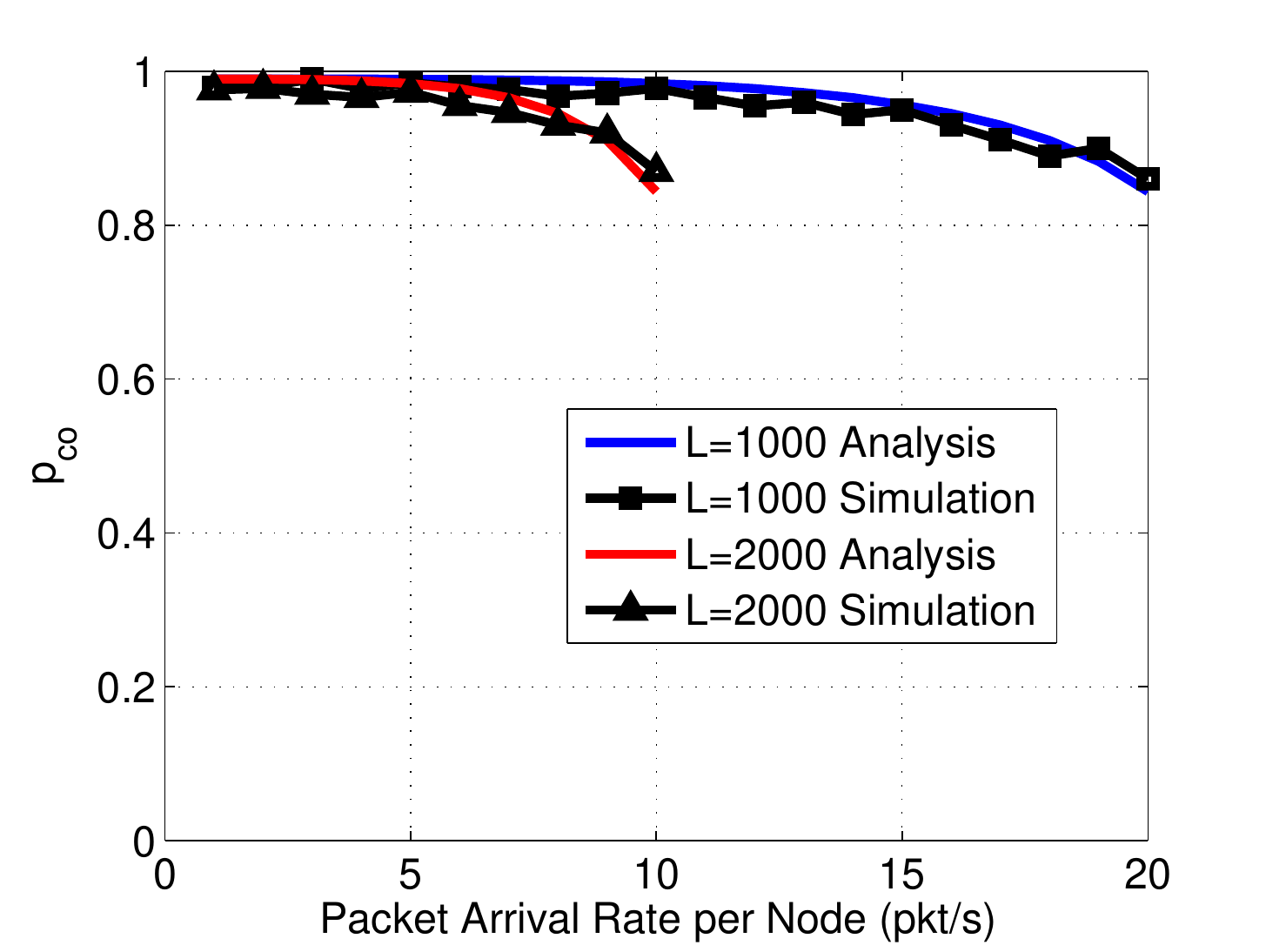}\label{fig:sg_load}}\hfill
\subfloat[$p_{co}$ vs. $n$ ($\lambda=10$), single-hop.]
    {\includegraphics[trim=4mm 1mm 11mm 7mm,clip,width=\twofig\linewidth]{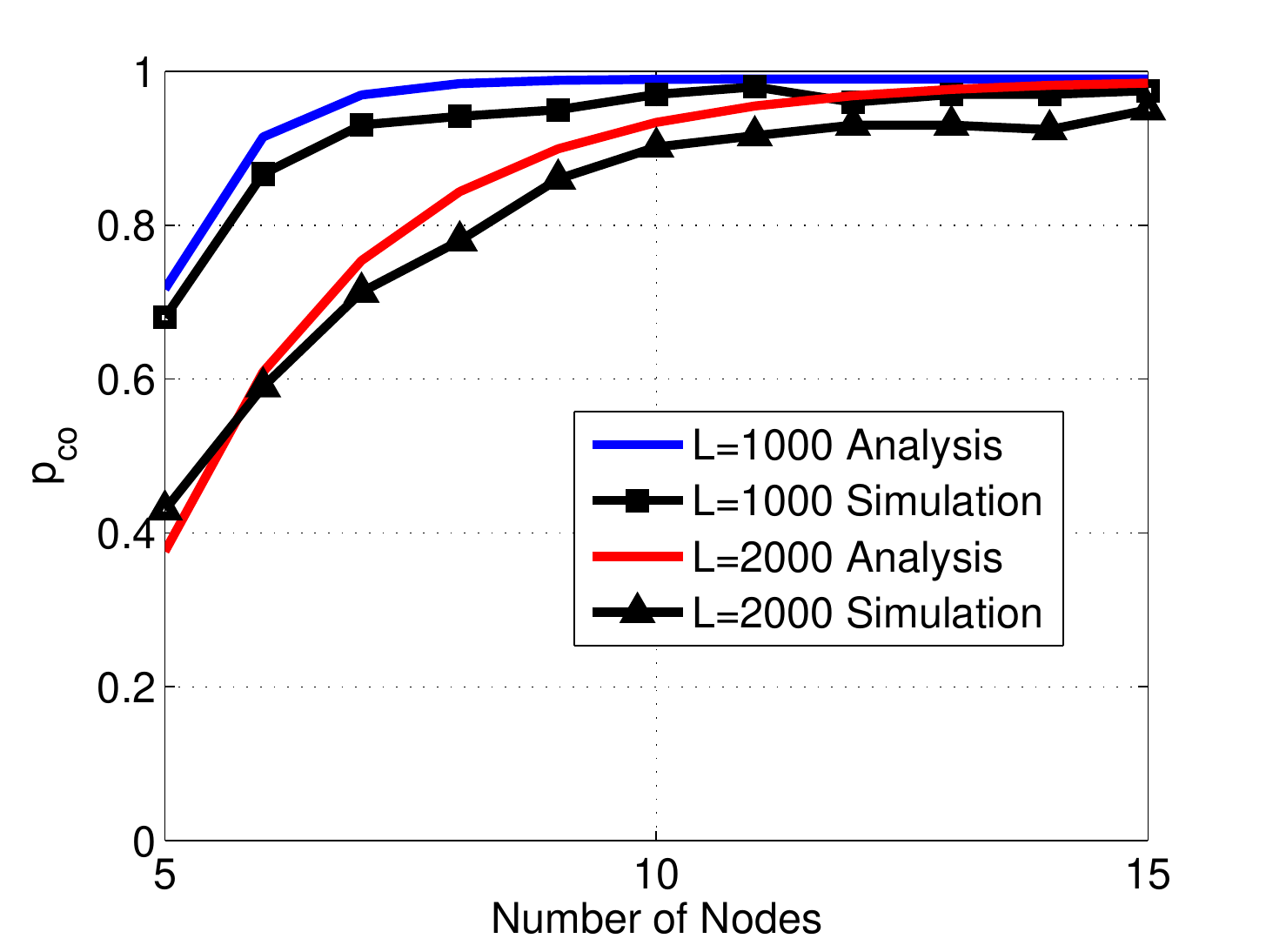}\label{fig:sg_node}}}
\centerline{
\subfloat[$p_{co}$ vs. $\lambda$ ($n=10$), multi-hop.]
    {\includegraphics[trim=4mm 1mm 11mm 7mm,clip,width=\twofig\linewidth]{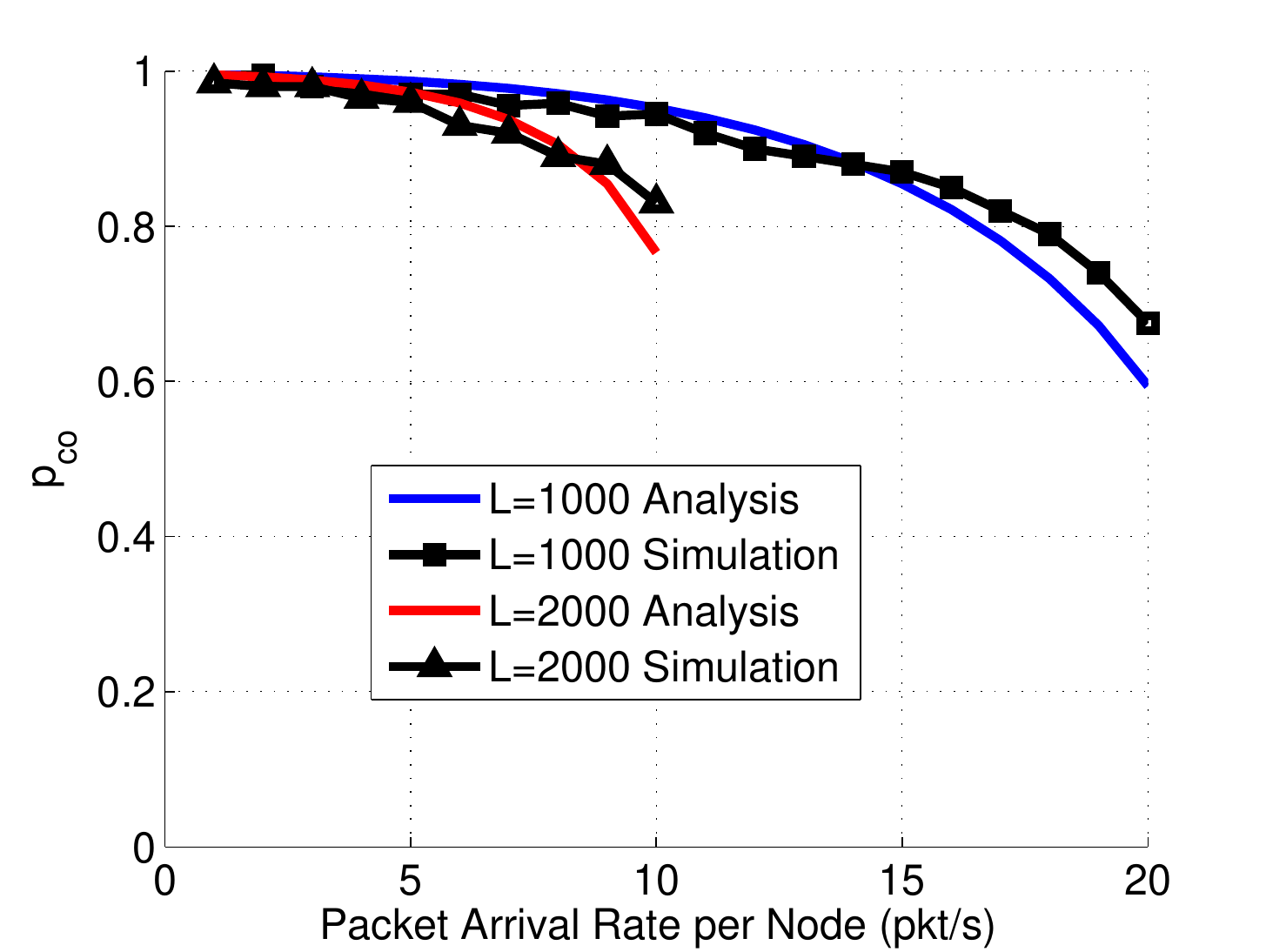}\label{fig:mt_load-pco}}\hfill
\subfloat[$p_{co}$ vs. $n$ ($\lambda=10$), multi-hop.]
    {\includegraphics[trim=4mm 1mm 11mm 7mm,clip,width=\twofig\linewidth]{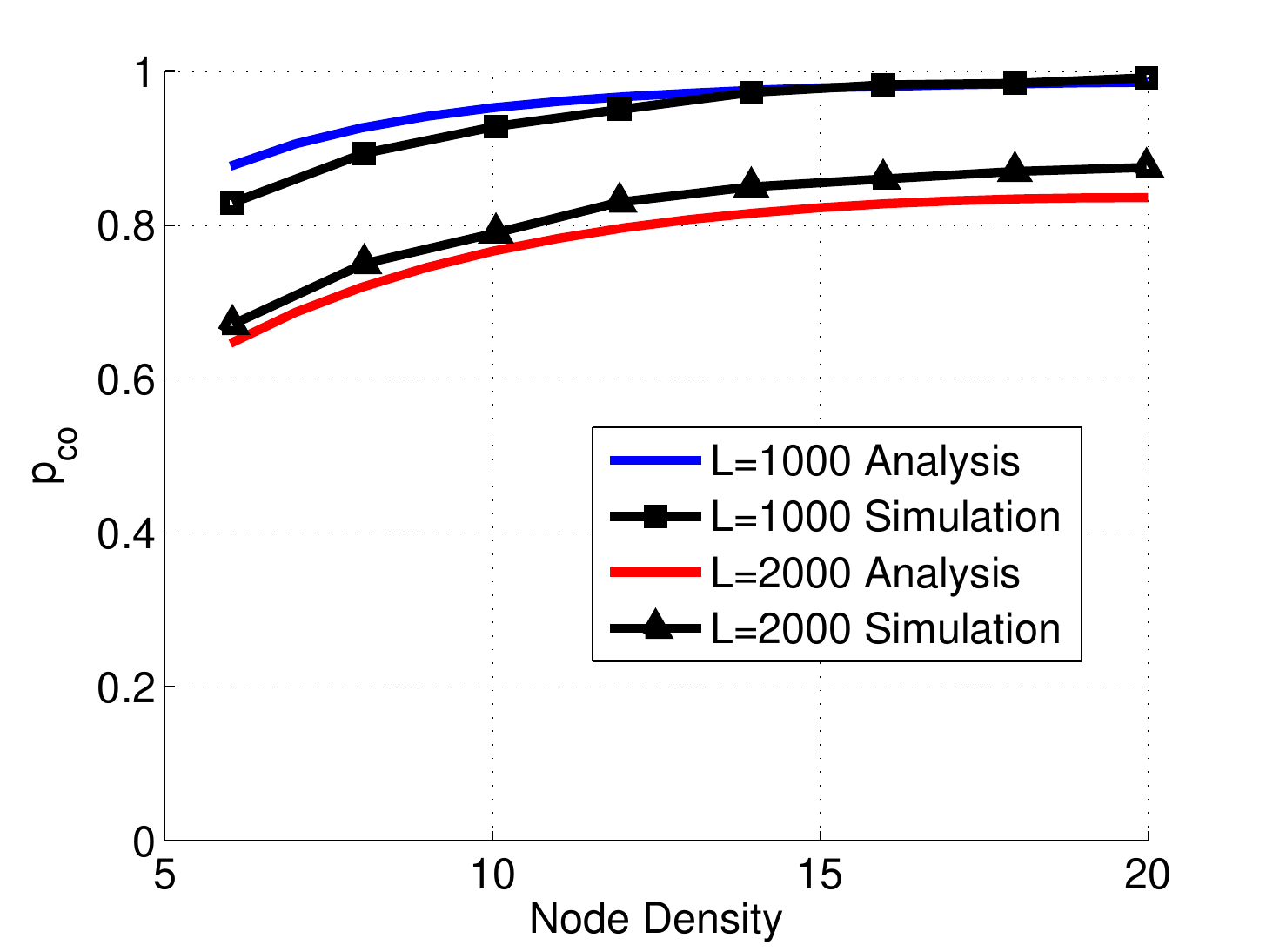}\label{fig:mt_node-pco}}}
\caption{Non-cooperative case: Impact of traffic load and node density, with different packet sizes. The value ranges of X axes are chosen such that the network is stable.}
\label{fig:result}
\end{figure}

\subsubsection{Verification of Analysis}

The $p_{co}$ obtained via analysis and simulations are compared in \fref{fig:result}. We see a close match between them, with a deviation of less than 5\% in almost all single-hop scenarios, 
and less than 10\% in almost all multi-hop scenarios. Particularly, the availability of cooperation is observed to be at a high level ($p_{co}>0.7$ in most cases), which suggests that a large percentage of MCC problems would be avoided by exploiting DISH, and DISH is feasible to use in multi-channel MAC protocols. ({\bf Finding 1})

Specifically, \fref{fig:sg_load} and \fref{fig:mt_load-pco} consistently show that, in both single-hop and multi-hop networks, $p_{co}$ monotonically decreases as $\lambda$ increases. The reasons are two folds. First, as traffic grows, each node spends more time on data channels for data transmission and reception, which reduces $p_{ctrl}$ and hence the chance of overhearing control messages ($p_{oh}$), resulting in lower $p_{co}$. Second, as the control channel is the rendezvous to set up all communications, larger traffic intensifies the contention and introduces more interference to the control channel, which is hostile to messages overhearing and thus also reduces $p_{co}$.

\fref{fig:sg_node} and \fref{fig:mt_node-pco} show that $p_{co}$ monotonically increases as $n$ increases, and is concave. The increase of $p_{co}$ is because MCC problems are more likely to have cooperative nodes under a larger node population, while the deceleration of the increase is because more nodes also generate more interference to the control channel.

An important message conveyed by this observation is that, although a larger node density creates more MCC problems (e.g., more channel conflicts as data channels are more likely to be busy), it also boosts the availability of cooperation which avoids more MCC problems. This implies that the performance degradation can be mitigated. ({\bf Finding 2})

In both single-hop and multi-hop networks, a larger packet size $L$ corresponds to a lower $p_{co}$. However, note that this is observed under the same {\em packet} arrival rate (pkt/s), which means actually a larger {\em bit} arrival rate for a larger $L$, and can be explained by the previous scenarios of $p_{co}$ versus $\lambda$. Now if we consider the same {\em bit} arrival rate, by examining the two analysis curves in \fref{fig:mt_load-pco} where we compare $p_{co}$ with respect to the same $\lambda\cdot L$ product, e.g., ($\lambda=5,L=2000$) versus ($\lambda=10,L=1000$), and ($\lambda=10,L=2000$) versus ($\lambda=20,L=1000$), then we will see that a larger $L$ corresponds to a {\em higher} $p_{co}$, which is contrary to the observation under the same {\em packet} arrival rate. The explanation is that, for a given bit arrival rate, increasing $L$ reduces the number of packets and hence {\em fewer control channel handshakes} are required, thereby alleviating control channel interference. ({\bf Finding 3})

\subsubsection{Dominating Impact Factor}\label{sec:domin}

\begin{figure}[tb]
\centering
\ifdefined\thesis
\includegraphics[trim=1cm 7mm 1cm 13mm,clip,width=0.7\linewidth]{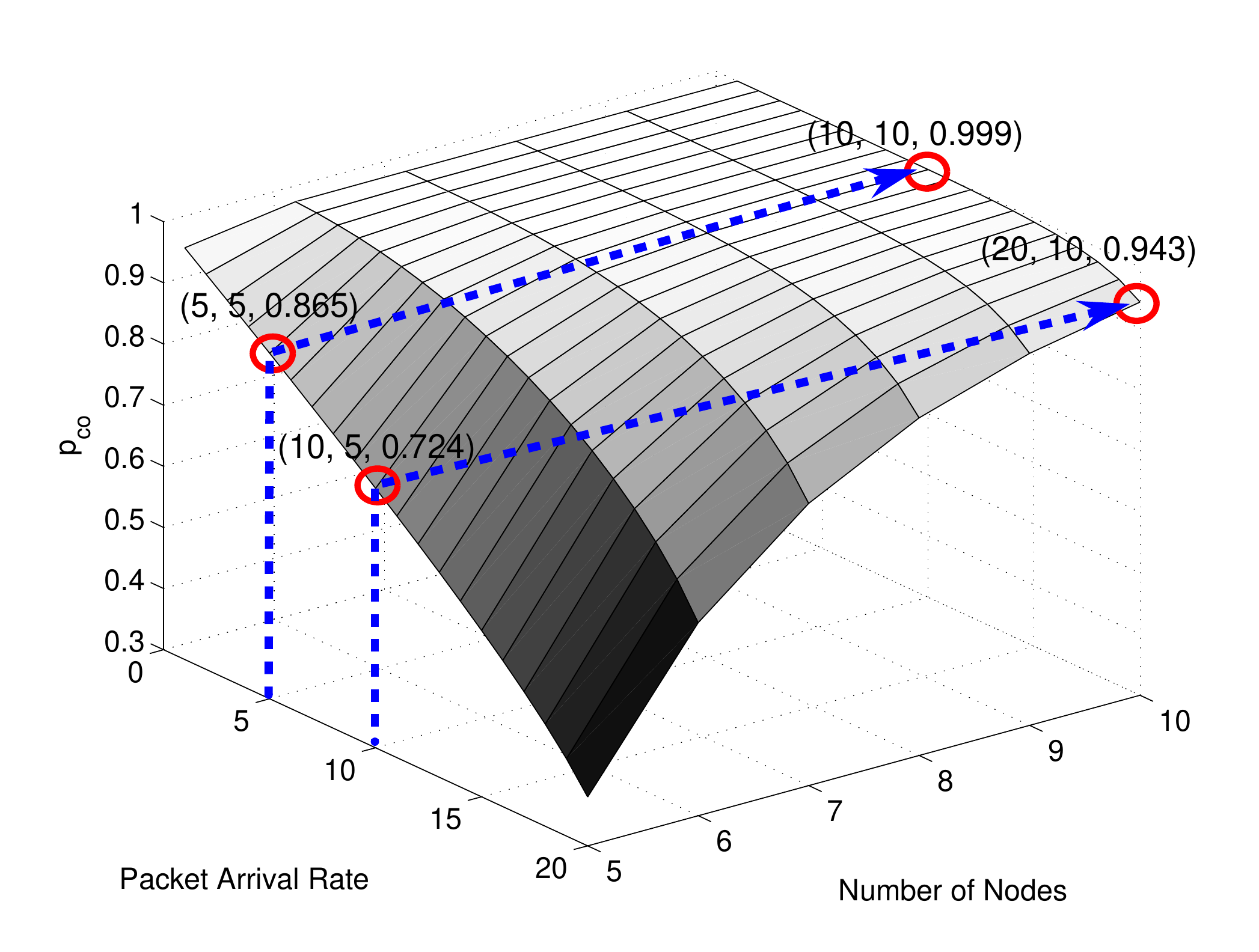}
\else
\includegraphics[trim=1cm 7mm 1cm 13mm,clip,width=0.8\linewidth]{3d_sg_L1k}
\fi
\caption{$p_{co}$ versus $\lambda$ and $n$. Each of the two arrows indicates a multiplicative increase of $\lambda$ and $n$ with the same factor (two).}
\label{fig:3d_sg_L1k}
\end{figure}

The above results indicate that node density and traffic load affect the availability of cooperation in {\em opposite} ways. This section aims to find which one dominates over the other.
In \fref{fig:3d_sg_L1k}, we plot the relationship of $p_{co}$ versus $\lambda$ and $n$, given $L=1000$ and based on the {\em analytical} result for single-hop networks. We multiplicatively increase $\lambda$ and $n$ with the same factor (two), and find that, when increasing ($\lambda,n$) from (5,5) to (10,10), $p_{co}$ keeps {\em increasing} from 0.865 to 0.999, and when increasing ($\lambda,n$) from (10,5) to (20,10), $p_{co}$ keeps {\em increasing} from 0.724 to 0.943.
Consistent results were also observed in other scenarios, i.e., $L=2000$ in single-hop, and $L=1000,2000$ in multi-hop networks.

This investigation shows that $n$ is the dominating factor over $\lambda$ that determines the variation of $p_{co}$. This implies that DISH networks should have better scalability than non-DISH networks, since $p_{co}$ increases when both traffic load and node density scale up. ({\bf Finding 4})

\subsection{Investigation with Ideal DISH}\label{sec:ideal}

For ideal DISH, we only present results in multi-hop networks, as the single-hop simulation results were similar and the discussion in this section applies to both sets of results.

\ifdefined\thesis
\begin{figure}[tb]
\centerline{
\subfloat[Impact of $\lambda$ ($n=10$).]
    {\includegraphics[trim=11mm 1mm 1cm 7mm,clip,width=0.45\linewidth]{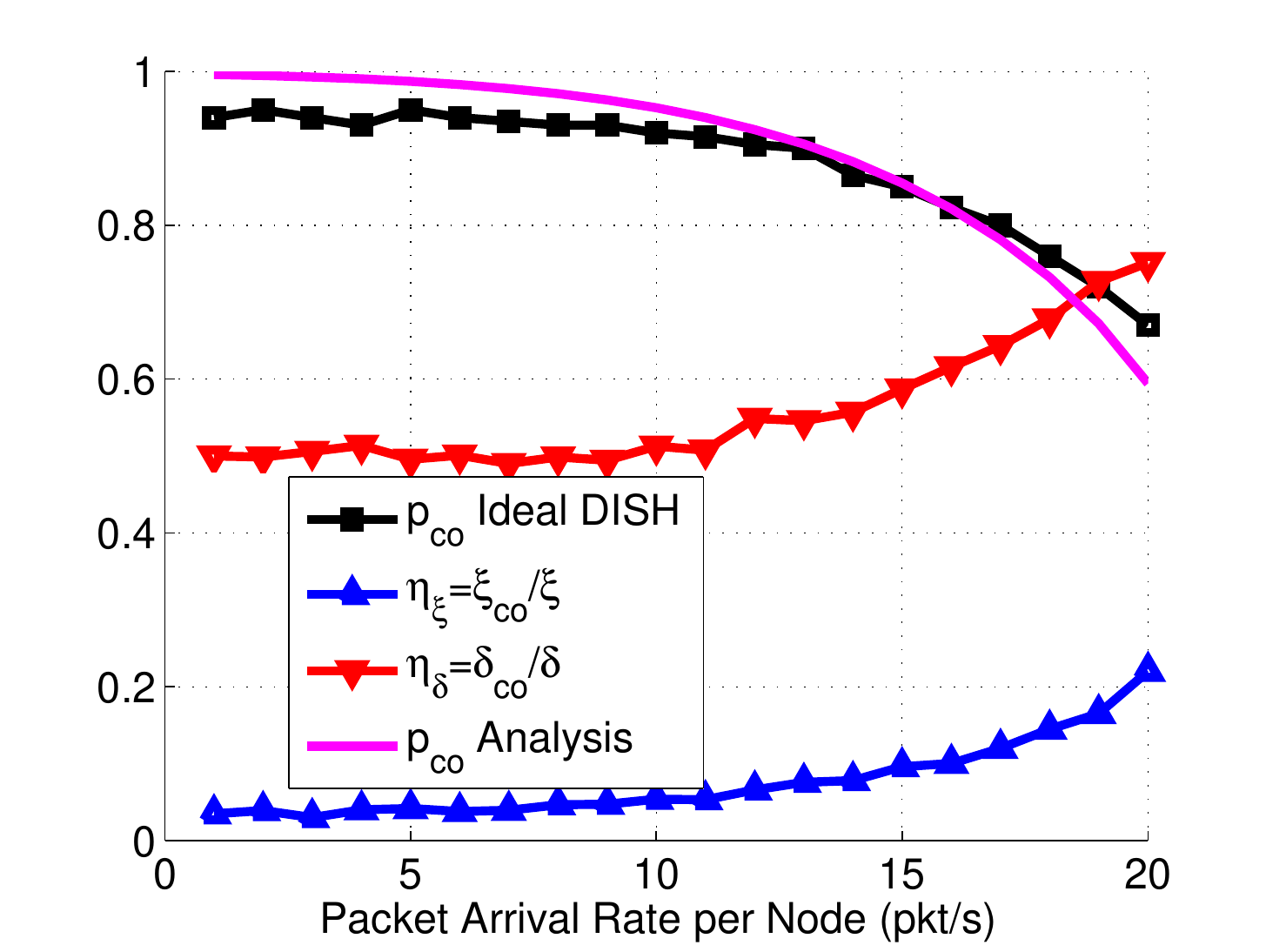}\label{fig:stable_ratio_load}}
\subfloat[Impact of $n$ ($\lambda=10$).]
    {\includegraphics[trim=11mm 1mm 1cm 7mm,clip,width=0.45\linewidth]{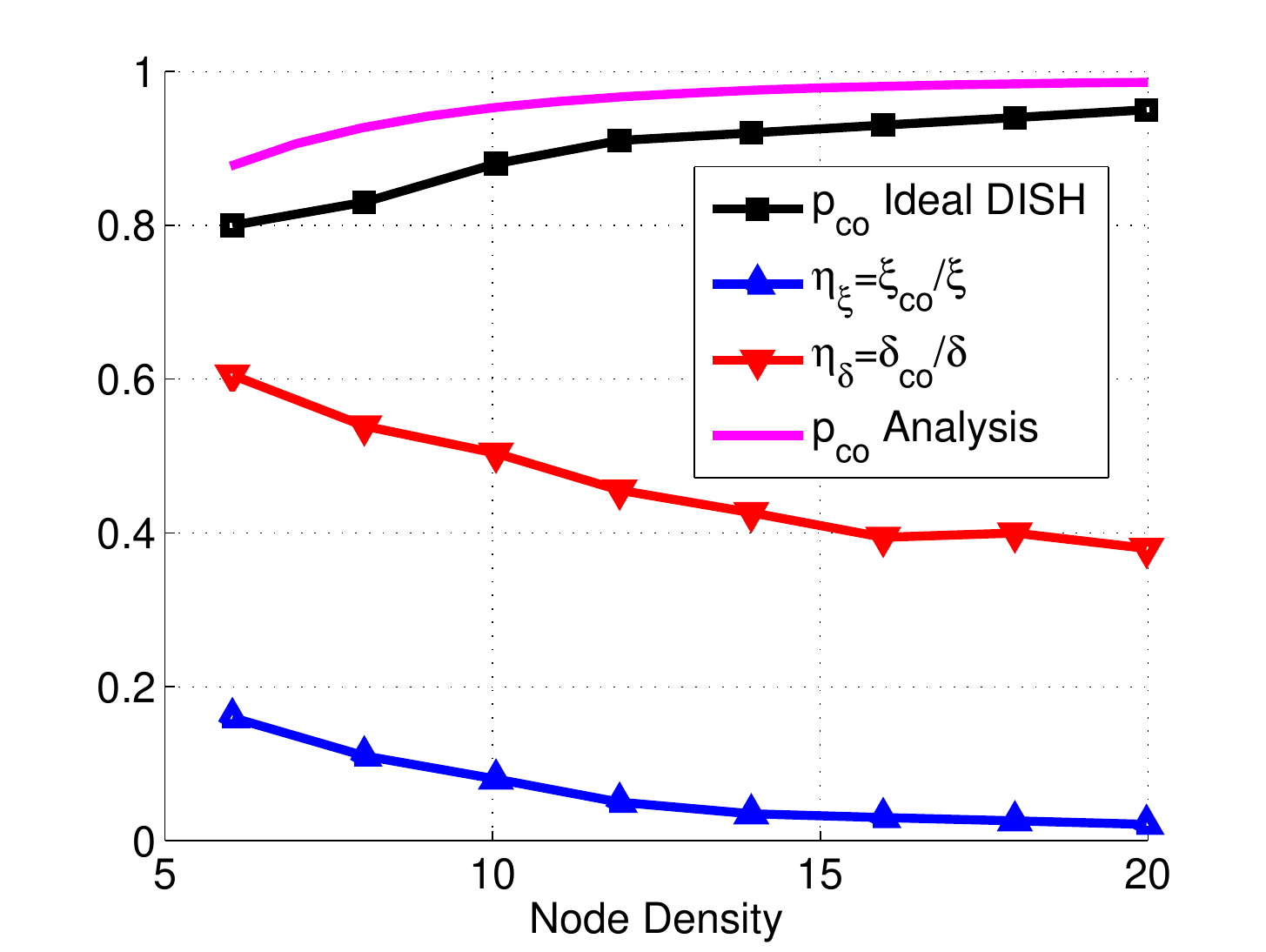}\label{fig:stable_ratio_node}}}
\caption{Investigating $p_{co}$ with ideal DISH in stable networks. This includes (i) verification of analysis, and (ii) correlation between $p_{co}$ and ($\eta_\xi, \eta_\delta$) (ratio of data collision, ratio of packet delay). $L=1000$ bytes. Each Y axis represents multiple metrics.}
\label{fig:stable}
\end{figure}
\else
\begin{figure*}[tb]
\begin{minipage}[tb]{0.6\linewidth}
\subfloat[Impact of $\lambda$ ($n=10$).]
    {\includegraphics[trim=11mm 1mm 1cm 7mm,clip,width=0.48\linewidth]{stable_ratio_load_jnl}\label{fig:stable_ratio_load}}
\subfloat[Impact of $n$ ($\lambda=10$).]
    {\includegraphics[trim=11mm 1mm 1cm 7mm,clip,width=0.48\linewidth]{stable_ratio_node_jnl}\label{fig:stable_ratio_node}}
\caption{Investigating $p_{co}$ with ideal DISH in stable networks. This includes (i) verification of analysis, and (ii) correlation between $p_{co}$ and ($\eta_\xi, \eta_\delta$) (ratio of data collision, ratio of packet delay). $L=1000$ bytes. The Y axis represents multiple metrics.}
\label{fig:stable}
\end{minipage}\hfil
\begin{minipage}[tb]{0.33\linewidth}
\rightline{
\includegraphics[trim=4mm 1mm 6mm 6mm,clip,width=.98\linewidth]{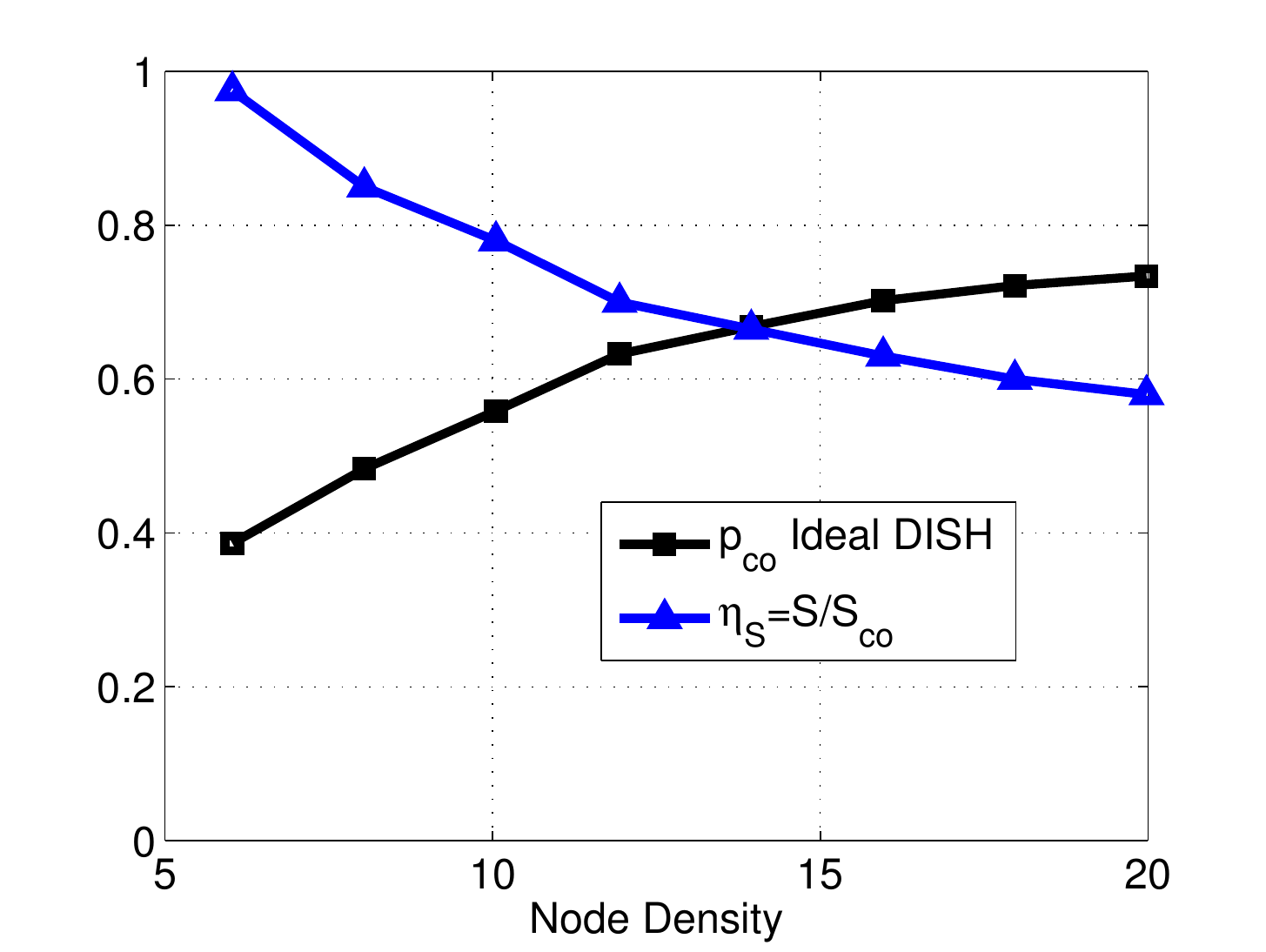}}
\caption{Investigating $p_{co}$ with ideal DISH in saturated networks: correlation between $p_{co}$ and $\eta_S$ (throughput ratio). $L=1000$ bytes. The Y axis represents multiple metrics.}
\label{fig:satu_ratio}
\end{minipage}
\end{figure*}
\fi

\subsubsection{Verification of Analysis}

The results of comparison are shown in \fref{fig:stable}, where $p_{co}$ with ideal DISH well matches $p_{co}$ of analysis. This confirms {\bf Findings 1-3}, and we speculate the reasons to be as follows. With ideal DISH, a transmitter will be informed of a deaf terminal problem at times and thus back off for a fairly long time, which leads to {\em fewer} {\tt McRTS} being sent. On the other hand, a node will also be informed of a channel conflict problem at times and thus re-select channel and retry shortly, which leads to {\em more} control messages being sent. Empirically, channel conflict problems occur more often, and hence there will be an overall {\em increase} of control messages being sent. This escalates interference and thus would lower down $p_{co}$. However, nodes will switch to data channels less frequently because cooperation suppresses a number of conflicting data channel usages. This makes nodes stay {\em longer} on the control channel and would elevate $p_{co}$. Consequently, $p_{co}$ does not significantly change.

As {\bf Finding 4} is obtained via analysis which, as shown in \fref{fig:stable}, matches the simulations with ideal DISH, it is automatically confirmed.

\subsubsection{Correlation between $p_{co}$ and Performance}

We investigate how $p_{co}$ correlates to network performance --- specifically, data channel collision rate $\xi$, packet delay $\delta$, and aggregate throughput $S$.
We consider both stable networks and saturated networks under multi-hop scenarios.

In stable networks, we measure ($\xi,\delta$) and $(\xi_{co},\delta_{co})$ when without and with cooperation (ideal DISH), respectively. Then we compute $\eta_\xi=\xi_{co}/\xi$ and $\eta_\delta=\delta_{co}/\delta$ to compare to $p_{co}$ with ideal DISH.  The first set of results, by varying traffic load $\lambda$, is shown in \fref{fig:stable_ratio_load}. %
We observe that the two {\em ascending} and {\em convex} curves of $\eta_\xi$ and $\eta_\delta$ approximately {\em reflect} the {\em descending} and {\em concave} curve of $p_{co}$, which hints at a {\em linear} or {\em near-linear} relationship between $p_{co}$ and these two performance ratios. That is, $\eta_\xi+p_{co}\approx c_1,\ \eta_\delta+p_{co}\approx c_2$, where $c_1$ and $c_2$ are two constants. The second set of results, by varying node density $n$, is shown in \fref{fig:stable_ratio_node}. %
On the one hand, $\eta_\xi$ and $\eta_\delta$ decreases as $n$ increases, which is contrary to \fref{fig:stable_ratio_load}. This confirms our earlier observations: $n$ is amicable whereas $\lambda$ is hostile to $p_{co}$ (the smaller $\eta_\xi$ and $\eta_\delta$, the better performance cooperation offers). On the other hand, the correlation between $p_{co}$ and the performance ratios is found again: as $p_{co}$ increases on a concave curve, it is reflected by $\eta_\xi$ and $\eta_\delta$ which decrease on two
convex curves.

In saturated networks, we vary node density $n$ and measure aggregate throughput without and with cooperation (ideal DISH), as $S$ and $S_{co}$, respectively. Then we compute $\eta_S=S/S_{co}$ (note that this definition is inverse to $\eta_\xi$ and $\eta_\delta$, in order for $\eta_S\in[0,1]$) to compare to $p_{co}$ with ideal DISH. The results are summarized in \fref{fig:satu_ratio}.
We see that (i) $p_{co}$ grows with $n$, which conforms to Finding 2, and particularly, (ii) the declining and convex curve of $\eta_S$ reflects the rising and concave curve of $p_{co}$, which is consistent with the observation in stable networks. In addition, the $p_{co}$ here is lower than the $p_{co}$ in stable networks. This is explained by our earlier result that higher traffic load suppresses $p_{co}$.

\ifdefined\thesis
\begin{figure}[tb]
\centering
\includegraphics[trim=4mm 1mm 6mm 6mm,clip,width=0.5\linewidth]{satu_ratio}
\caption{Investigating $p_{co}$ with ideal DISH in saturated networks: correlation between $p_{co}$ and $\eta_S$ (throughput ratio). $L=1000$ bytes. The Y axis represents multiple metrics.}
\label{fig:satu_ratio}
\end{figure}
\fi

In summary, the experiments in stable networks and saturated networks both demonstrate a strong correlation ({\em linear} or {\em near-linear} mapping) between $p_{co}$ and network performance ratio in terms of typical performance metrics. This may significantly simplify performance analysis for cooperative networks via bridging the {\em nonlinear} gap between network parameters and $p_{co}$, and also suggests that $p_{co}$ be used as an appropriate performance indicator itself. ({\bf Finding 5})

Note that this does not imply that the delay, the channel collision rate, and the throughput are linear with respect to each other, because the above stated relationship is for the performance {\em ratios} between DISH and non-DISH networks.

The explanation to this linear or near-linear relationship should involve intricate network dynamics. We speculate that the rationale might be that (i) MCC problems are an essential performance {\em bottleneck} to multi-channel MAC performance, and (ii) $p_{co}$ is equivalent to the {\em ratio} of MCC problems that can be avoided by DISH. In any case, we reckon that this observation may spur further studies and lead to more thought-provoking results.

\subsection{Investigation with Real DISH}\label{sec:cammac}

For the same reason as mentioned for ideal DISH, we present the results for multi-hop networks only.

\subsubsection{Verification of Analysis}

\begin{figure}[tb]
\centering
\ifdefined\thesis
\subfloat[$p_{co}$ versus $\lambda$ ($n=10$).]
    {\includegraphics[trim=4mm 0 11mm 7mm,clip,width=0.45\linewidth]{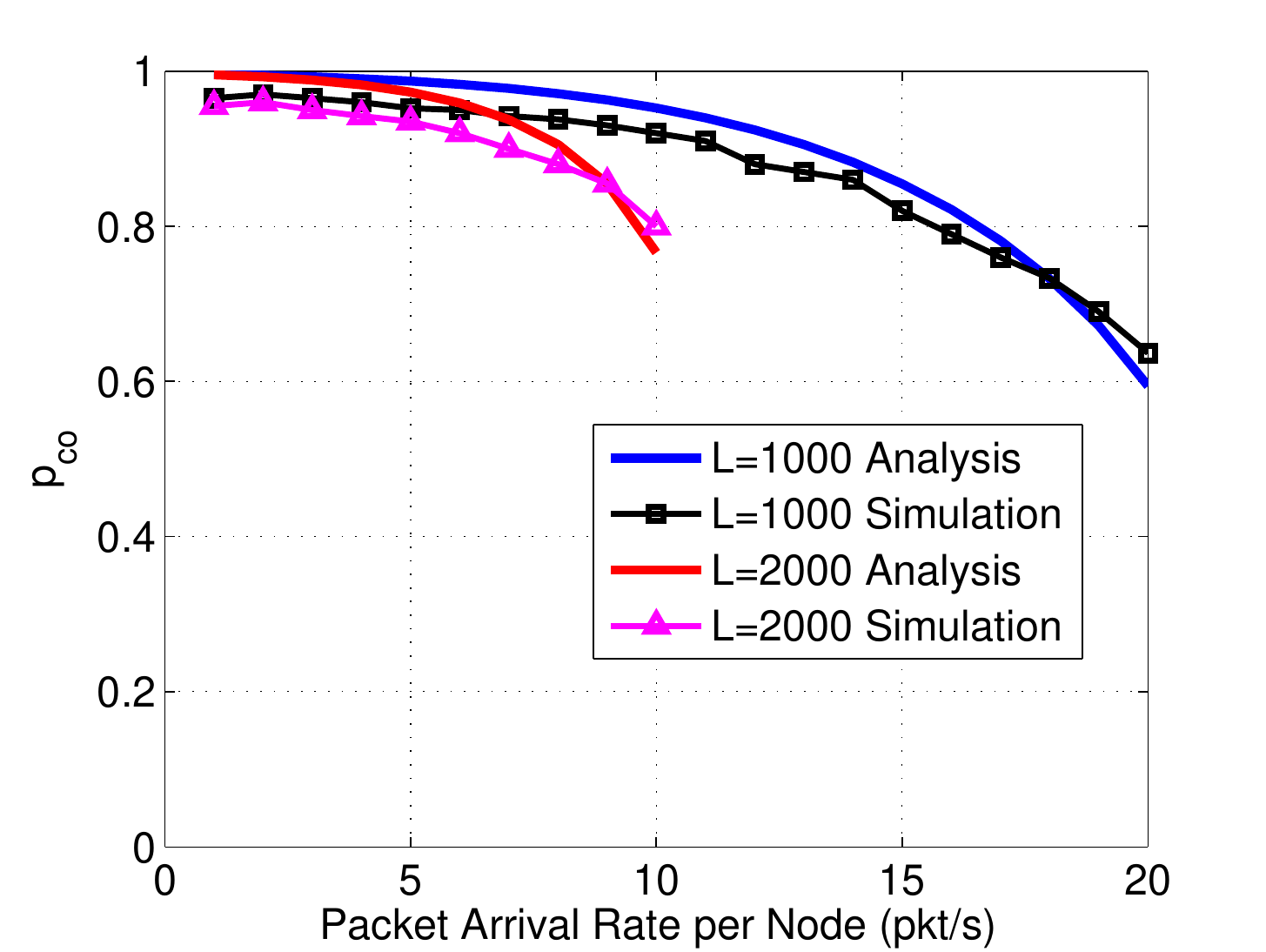}\label{fig:mt_pcocam_load}}
\subfloat[$p_{co}$ versus $n$ ($\lambda=10$).]
    {\includegraphics[trim=4mm 0 11mm 7mm,clip,width=0.45\linewidth]{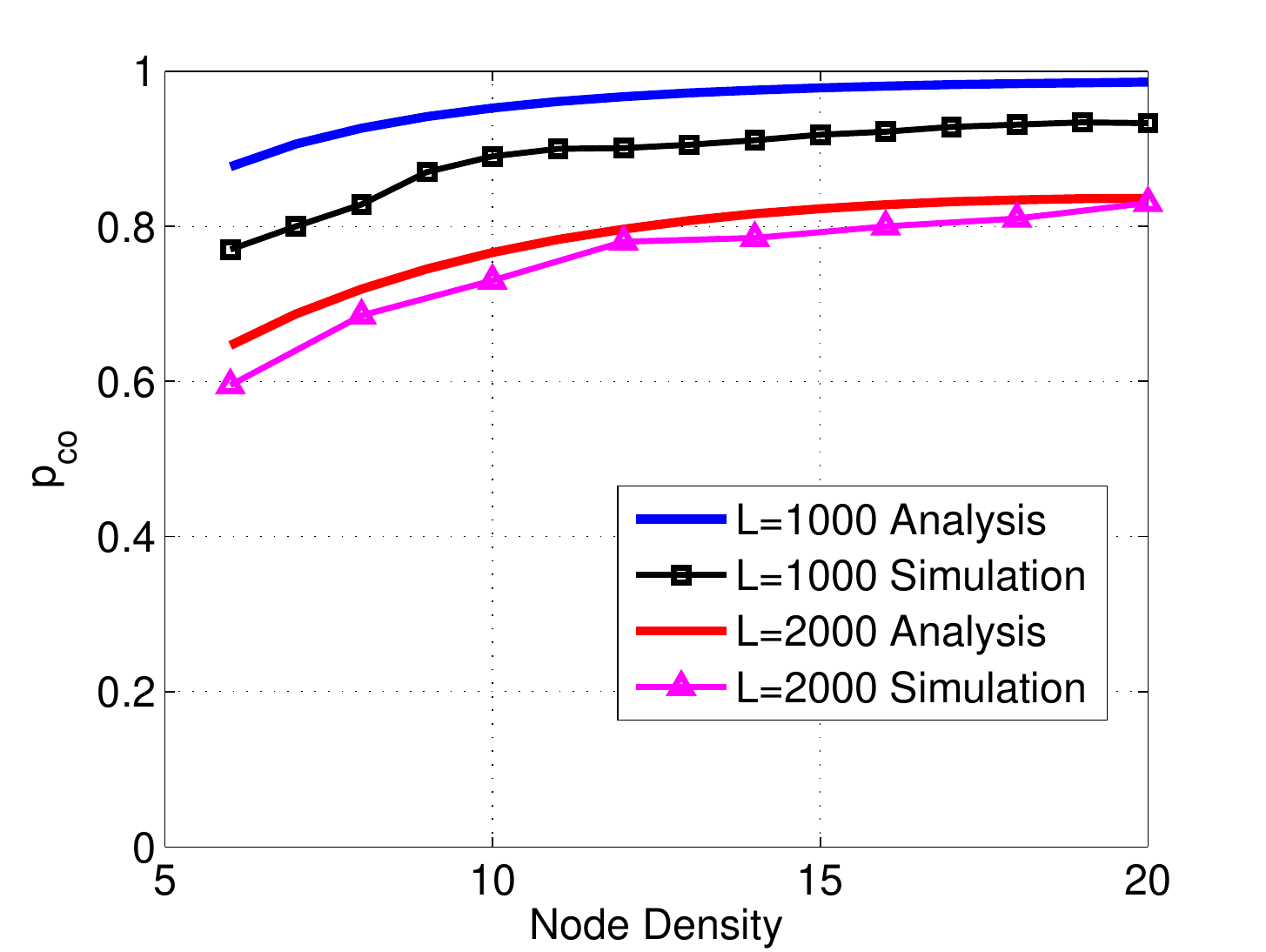}\label{fig:mt_pcocam_node}}
\else
\subfloat[$p_{co}$ versus $\lambda$ ($n=10$).]
    {\includegraphics[trim=3mm 0 11mm 7mm,clip,width=0.5\linewidth]{mt_pcocam_load}\label{fig:mt_pcocam_load}}
\subfloat[$p_{co}$ versus $n$ ($\lambda=10$).]
    {\includegraphics[trim=3mm 0 11mm 7mm,clip,width=0.5\linewidth]{mt_pcocam_node}\label{fig:mt_pcocam_node}}
\fi
\caption{Verification of $p_{co}$ with real DISH in multi-hop networks. $L=1000$ bytes.}
\label{fig:pcocam}
\end{figure}

As shown in \fref{fig:pcocam}, the simulation and analytical results still match, with deviation less than 15\%. This is explained by two underlying factors. On the one hand, since real DISH actually sends cooperative messages, the control channel will have more interference which tends to diminish $p_{co}$. On the other hand, these cooperative messages inform transmitters or receivers of conflicting channel selections, which leads to a reduced number of channel switchings. Hence nodes will stay longer on the control channel and hence $p_{co}$ tends to be elevated. Consequently, $p_{co}$ is kept close to the analytical result. {\bf Findings 1-4} are thus confirmed.

\subsubsection{Correlation between $p_{co}$ and Performance}

\begin{figure*}[tb]
\centerline{
\subfloat[$\eta_\delta$ and $\eta_\xi$ versus $p_{co}$, varying $\lambda$ ($n=10$).]
    {\includegraphics[trim=2mm 0 8mm 7mm,clip,width=0.33\linewidth]{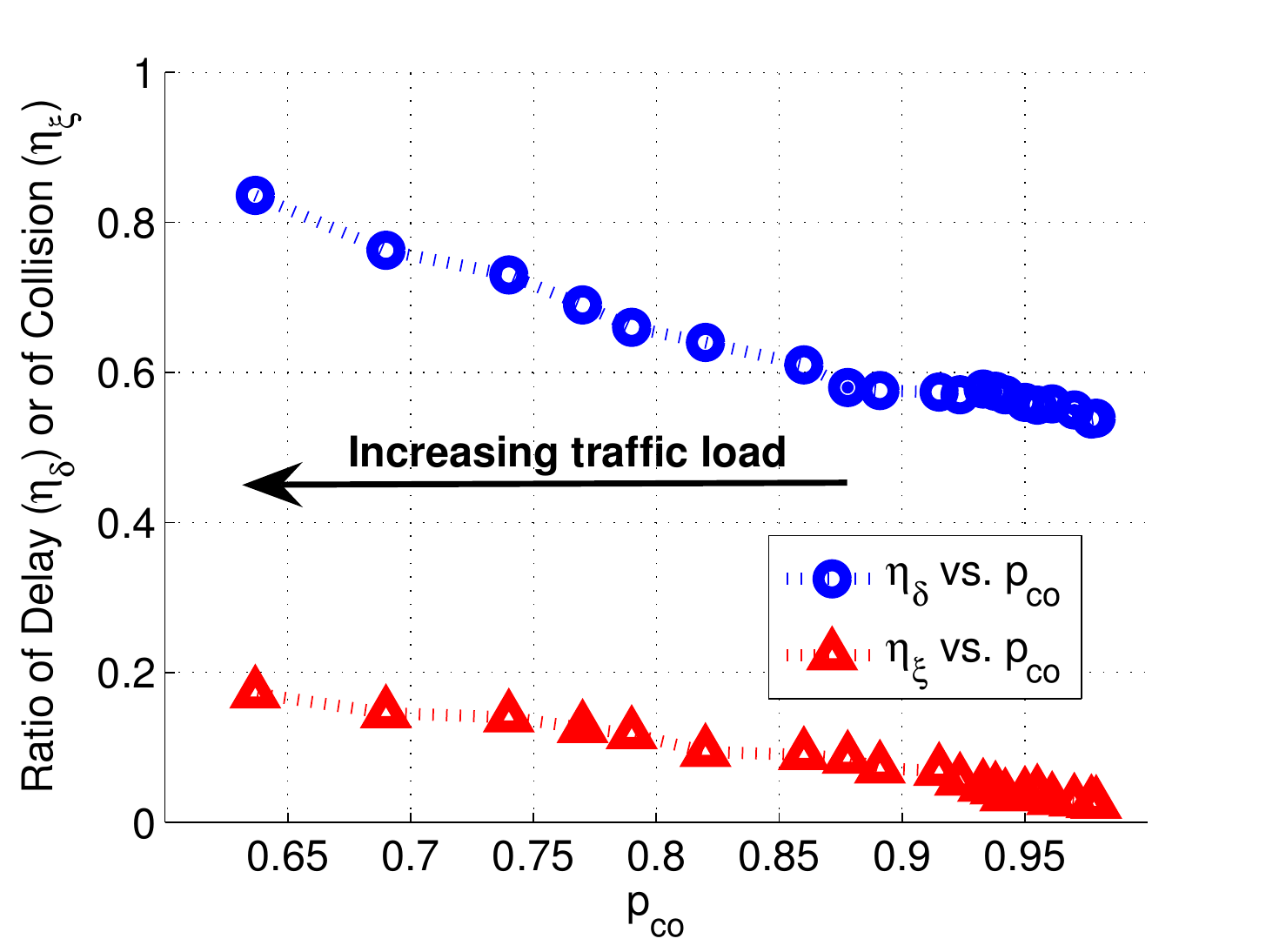}\label{fig:scatter_delaycoll_load}}
\subfloat[$\eta_\delta$ and $\eta_\xi$ versus $p_{co}$, varying $n$ ($\lambda=10$).]
    {\includegraphics[trim=2mm 0 9mm 7mm,clip,width=0.33\linewidth]{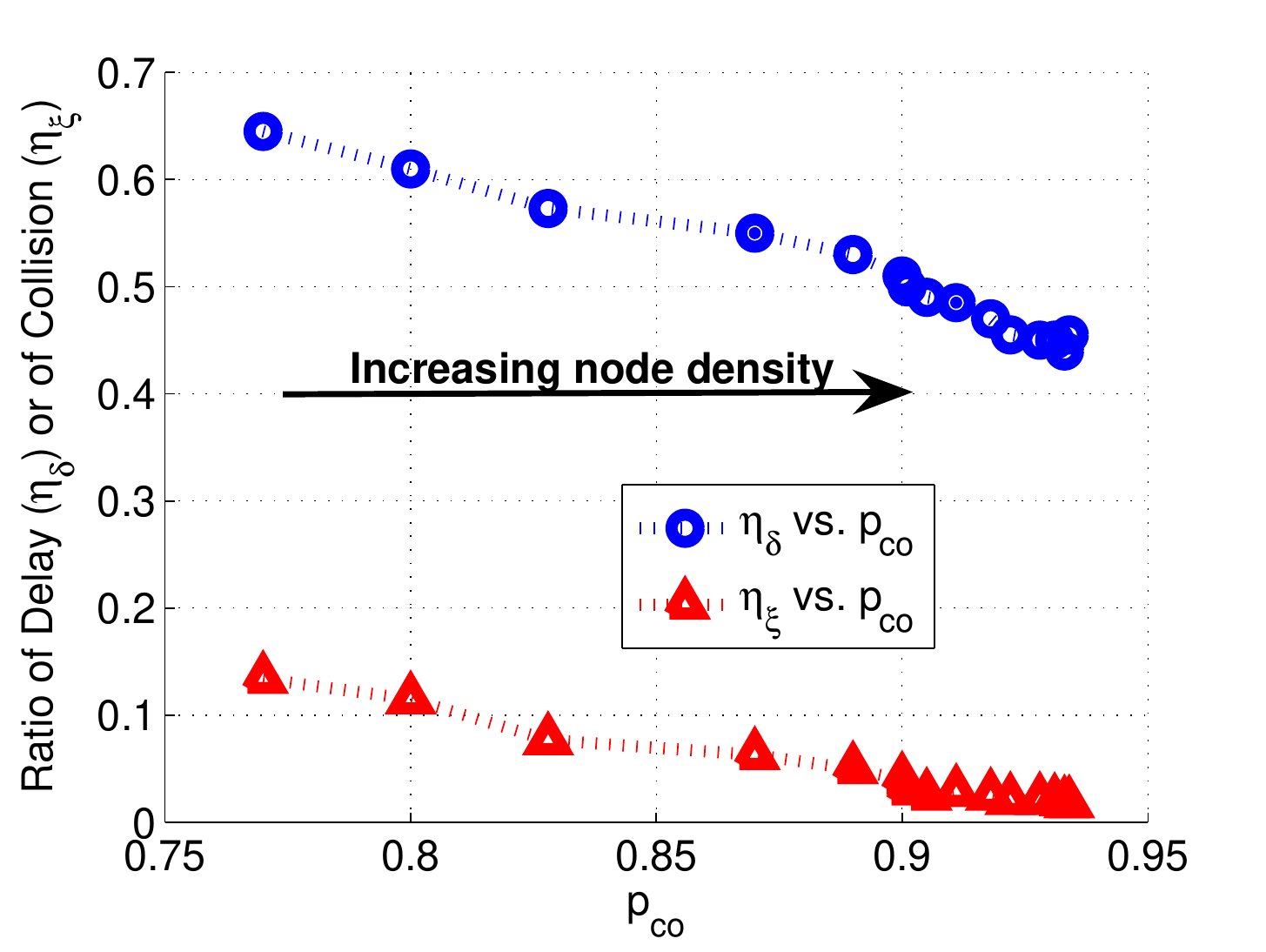}\label{fig:scatter_delaycoll_node}}
\subfloat[$\eta_S$ versus $p_{co}$.]
    {\includegraphics[trim=2mm 0 1cm 7mm,clip,width=0.33\linewidth]{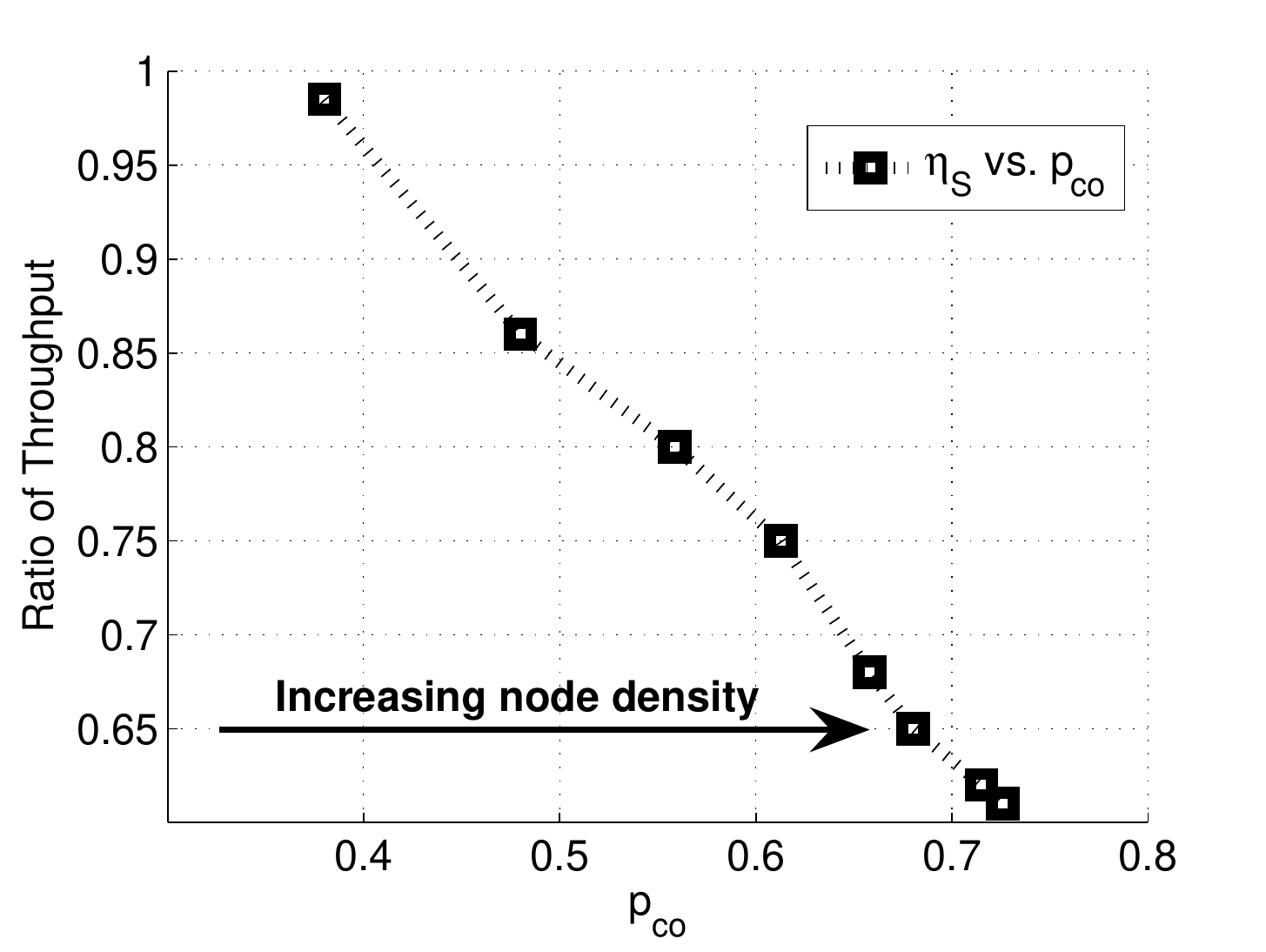}\label{fig:scatter_thpt}}}
\caption{Correlation between $p_{co}$ and different performance metrics with real DISH in multi-hop networks.}
\label{fig:scatter}
\end{figure*}

We examine this issue using scatter plots which provide another point of view besides the direct representation in \sref{sec:ideal}. These plots are given in \fref{fig:scatter}, where it is more apparent to see the near-linear relationship between $p_{co}$ and $\eta_\xi, \eta_\delta, \eta_S$ (respectively). This confirms {\bf Finding 5}. 

An observation is that, while $\eta_\delta$ and $\eta_S$ (the ratio of delay and throughput, respectively) in real DISH are slightly larger than those in ideal DISH, $\eta_\xi$ (the ratio of data channel collision) is slightly smaller. Note that, by definition, the smaller these ratios are, the better the corresponding performance is (i.e., shorter delay, fewer collisions, and higher throughput, respectively). To explain this difference, first notice that the larger $\eta_\delta$ and $\eta_S$ is simply because real DISH has to afford overhead for cooperation (physically send cooperative messages) while ideal DISH does not need to. Second, the smaller $\eta_\xi$, which counter-intuitively conveys fewer data channel collisions than ideal DISH, relates to two factors: (i) the transmissions of cooperative messages in real DISH lowers down the efficiency of cooperation than in ideal DISH, as explained before, and hence each use of a data channel in real DISH is slightly more likely to encounter collision, (ii) the cooperative messages suppress nearby nodes from {\em initiating} handshakes (via CSMA) and also interfere {\em ongoing} control channel handshakes, leading to fewer accomplished control channel handshakes per second, and hence fewer data channel usages than ideal DISH. The latter factor, according to the simulation results, outweigh the former factor, thereby explains the smaller $\eta_\xi$. In fact, these two factors also contribute to the longer delay and lower throughput in real DISH than in ideal DISH.

\section{Channel Bandwidth Allocation}\label{sec:bwalloc}

Our investigation shows that $p_{co}$ is a meaningful performance indicator for DISH networks and captures several other critical performance metrics. In this section, we leverage $p_{co}$ as an instrument and apply our analysis to solving an important issue in multi-channel operation, channel bandwidth allocation.

\subsection{Problem Formulation}

For a given amount of total channel bandwidth $W$, define a bandwidth allocation scheme $\A A_m=(m, \sigma)$, where $m$ is the number of data channels and $\sigma = w_c/w_d$, in which $w_c$ is the bandwidth of the control channel and $w_d$ is the bandwidth of a data channel. The objective is to obtain the optimal scheme $\A A_m^*=(m, \sigma^*)$ which achieves the maximum $p_{co}$ for a given $m$. We remark that:
\begin{itemize}
\item An implicit assumption of the problem formulation is that bandwidth is equally partitioned among all {\em data} channels.
\item In the formulation, $m$ is designated as an input parameter instead of a variable subject to optimization. This is because of the following: (i) in practice, the main consideration on choosing $m$ is {\em system capacity} (the number of users a system can accommodate),
(ii) if, otherwise, $m$ is a variable subject to optimization, the formulation will be equivalent to $\A A=(w_c, w_d)$ and its solution will be a {\em single} ``universally optimal'' $m$ which generally does not fit into practical situations.
\item There exists another bandwidth allocation problem, where data channel bandwidth $w_d$ is fixed and only control channel bandwidth $w_c$ can be adjusted (or vice versa). We do not investigate this problem because (i) from a practical perspective, radio frequency band is a regulated or highly limited resource and cannot be arbitrarily claimed, and (ii) even if the band can be arbitrarily claimed, then the solution becomes obvious --- $p_{co}$ will monotonically increase as $w_c$ or $w_d$ grows, as a consequence of using more resource.
\end{itemize}

In the following, we solve the above optimization problem using the analytical results derived in \sref{sec:analysis}. The feasibility of applying our analysis is based on the consistency between $p_{co}$ of analysis and that of simulations across {\red non-cooperative case}, ideal DISH and real DISH, as shown in \sref{sec:simu-pco}.

\subsection{Solutions and Discussion}

Denoting the size of a control packet and a data packet by $l_c$ and $L$, respectively, we have\footnote{We assume that ACK packet size is negligible compared to data packet size. Or alternatively, one can simply define $L$ as the sum size of data and ACK packets.}
\[ l_c = w_c b, \;\; L = w_d T_d. \]
Combined with $W = w_c + m \cdot w_d$, we have
\begin{align}\label{eq:totalbw}
\frac{l_c}{b} + \frac{m L}{T_d} = W.
\end{align}
Then we rewrite $\sigma$ as
\begin{align}\label{eq:sigma}
\sigma = \frac{T_d l_c}{b L}.
\end{align}

Next, for a given $W$ and protocol-specified $l_c$ and $L$, compute $b$ and $T_d$ using \eqref{eq:totalbw} and \eqref{eq:sigma} for different combinations of $m$ and $\sigma$. Then substitute each pair of $b$ and $T_d$ into the equations derived in \sref{sec:analysis} and calculate $p_{co}$. By this means, we will obtain a matrix of $p_{co}$ which corresponds to different combinations of $m$ and $\sigma$. Finally, comparing $p_{co}$ for each $m$ in order to find the maximum will obtain the optimal solution $\A A_m^*=(m, \sigma^*)$ for a given $m$.

\begin{figure}[tb]
\centering
\ifdefined\thesis
\includegraphics[trim=3mm 4mm 11mm 8mm, clip,width=0.65\linewidth]{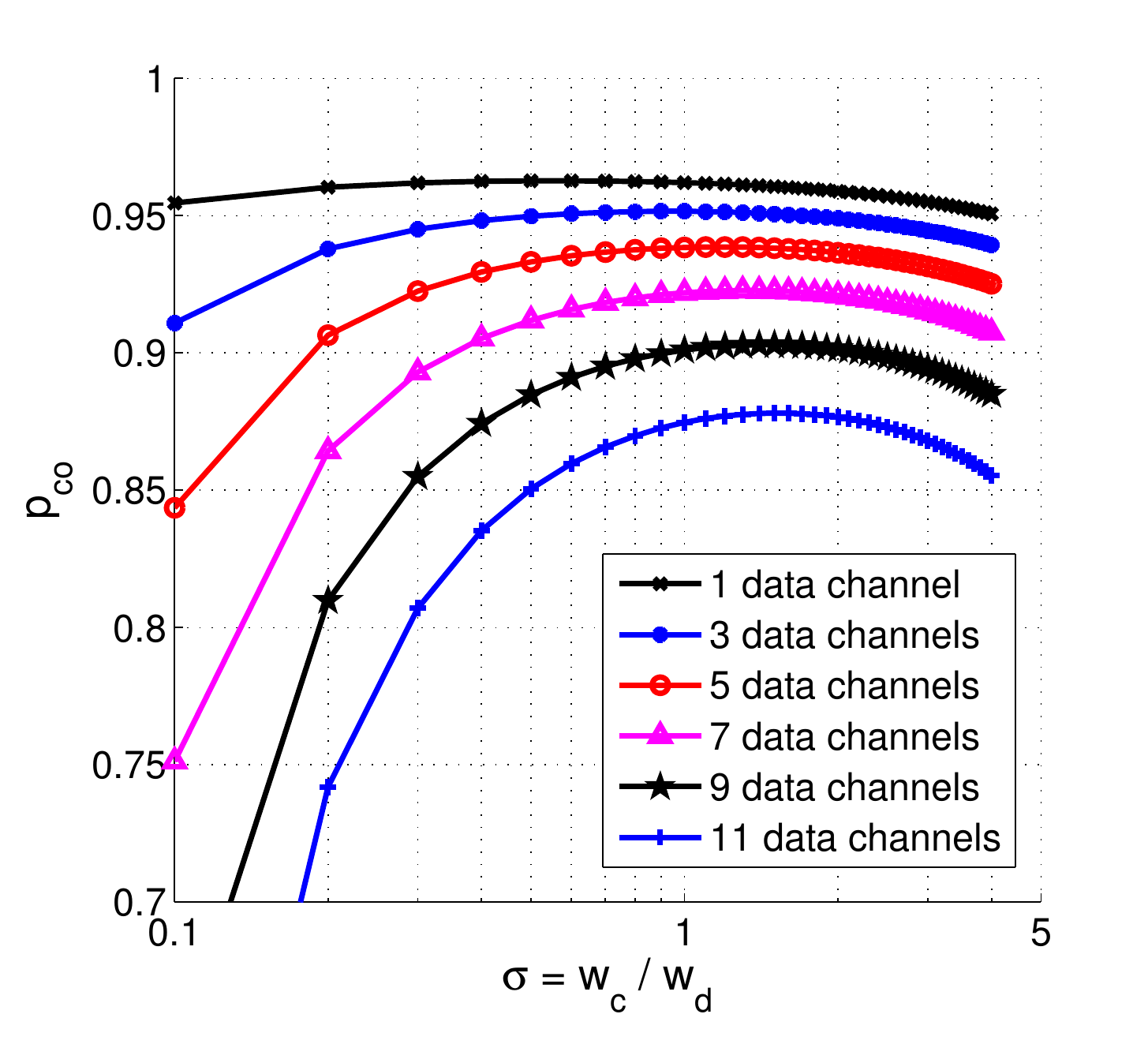}
\else
\includegraphics[trim=3mm 4mm 11mm 8mm, clip,width=0.75\linewidth]{bw_pco_sigma}
\fi
\caption{$p_{co}$ versus $\sigma$ under different $m$.}
\label{fig:bwalloc}
\vspace{-3mm}\end{figure}

Using this method, we can demonstrate results given a set of parameters. \fref{fig:bwalloc} is such a plot given the following parameters: $L=2000$ bytes, $l_c=34$ bytes (cf. \fref{fig:format-pco}, plus PHY preamble and header), $W=40$Mb/s, $n=6$, $\lambda=20$.  We can see that $p_{co}$ is concave and {\em not} monotonic with respect to $\sigma$, and it reaches the maximum at a certain $\sigma$ on each curve corresponding to a specific $m$. There are two counteractive factors attributing to this. First, as $\sigma$ increases, the control channel is allocated more bandwidth, so the time needed for transmitting a control packet, and hence the vulnerable period of receiving a control packet, is being reduced. As such, the probability of successful overhearing will increase and hence $p_{co}$ tends to be higher. Second, as $W$ is fixed, increasing control channel bandwidth has to squeeze data channels simultaneously, which prolongs data packet transmission time and hence is to the effect of enlarging data packet size $L$. According to \sref{sec:simu-pco}, a larger $L$ leads to a lower $p_{co}$. ({\bf Finding 6})

In particular, we obtain $\A A_1^*=(1, 0.55)$, $\A A_3^*=(3, 0.95)$, $\A A_5^*=(5, 1.15)$, $\A A_7^*=(7, 1.35)$, $\A A_9^*=(9, 1.45)$, $\A A_{11}^*=(11, 1.5)$. This conveys the following message: when the number of channels is small, the control channel should be allocated less bandwidth than a data channel (i.e., $\sigma<1$), whereas when there are more channels, the control channel should be allocated more bandwidth than a data channel (i.e., $\sigma>1$). The rationale is that, as the number of channels increases, it becomes easier for a node to secure a data channel for data transmission, and thus fewer nodes will be waiting for free data channels on the control channel. This reduces the chances of having cooperative nodes. In order to counteract this effect, the control channel should be allocated more bandwidth (than equal partition) to increase the probability of successful overhearing (by shortening the vulnerable periods of receiving control packets).

\begin{figure}[tb]
\centering
\ifdefined\thesis
\subfloat[Under different $n$ ($\lambda=10$).]
{\includegraphics[trim=5mm 2mm 11mm 6mm, clip,width=0.45\linewidth]{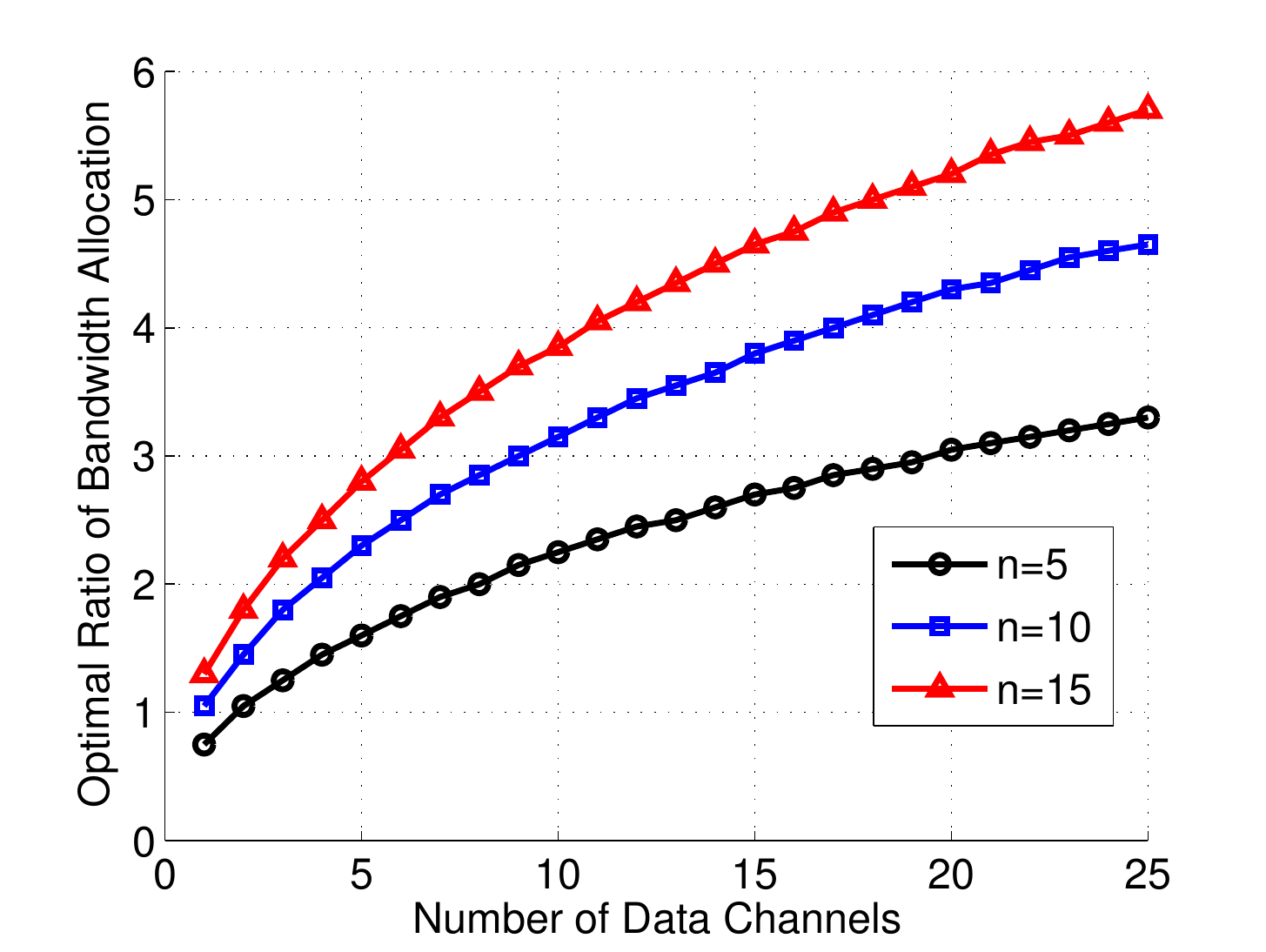}
    \label{fig:bw_optsig_diffn}}
\subfloat[Under different $\lambda$ ($n=6$).]
{\includegraphics[trim=5mm 2mm 11mm 7mm, clip,width=0.45\linewidth]{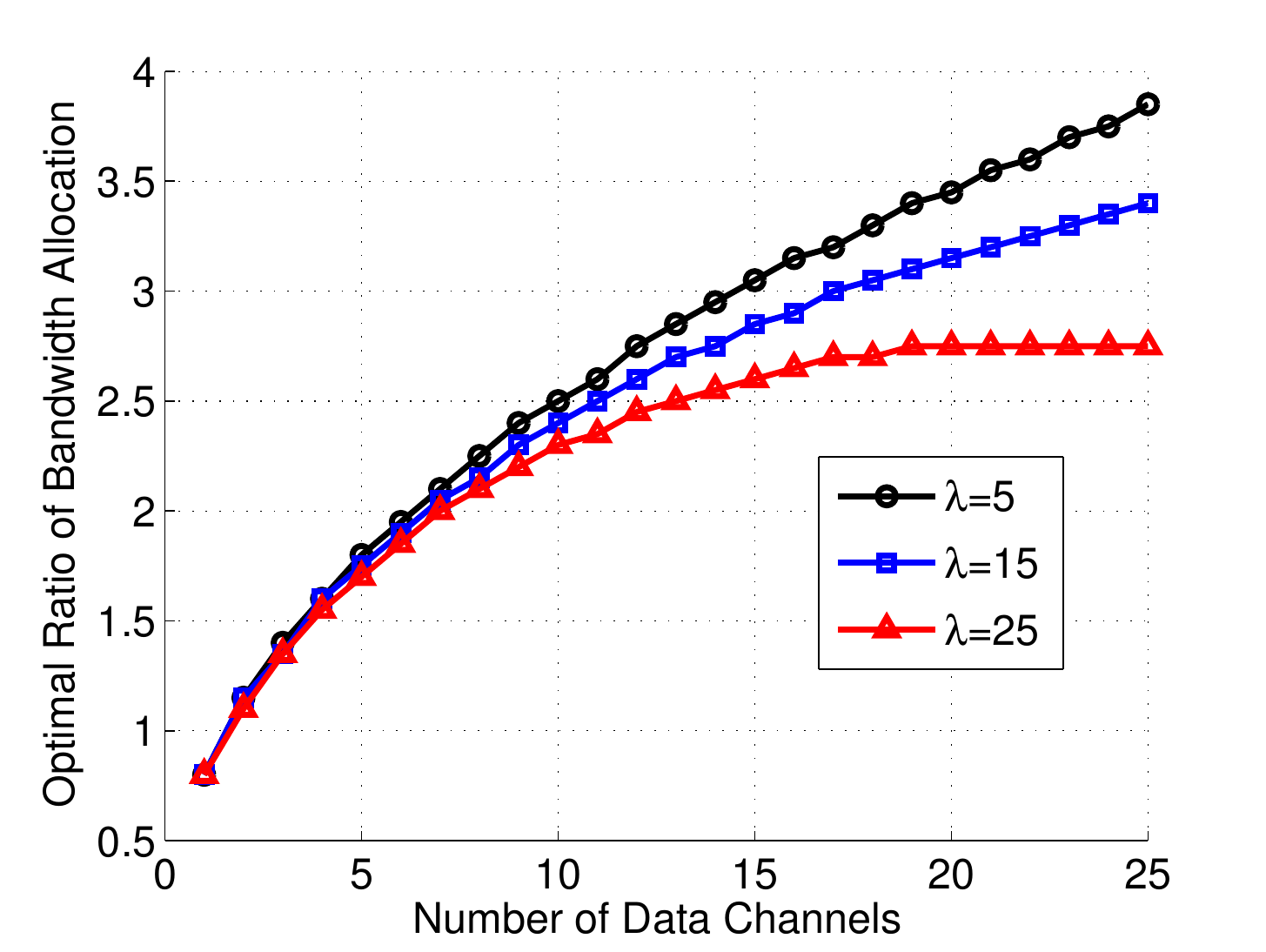}
    \label{fig:bw_optsig_difflmd}}
\else
\subfloat[Under different $n$ ($\lambda=10$).]
{\includegraphics[trim=5mm 2mm 11mm 6mm, clip,width=0.5\linewidth]{bw_optsig_diffn}
    \label{fig:bw_optsig_diffn}}
\subfloat[Under different $\lambda$ ($n=6$).]
{\includegraphics[trim=5mm 2mm 11mm 7mm, clip,width=0.5\linewidth]{bw_optsig_difflmd}
    \label{fig:bw_optsig_difflmd}}
\fi
\caption{$\sigma^*$ versus $m$ under different combinations of $n$ and $\lambda$. $L=1000$ bytes.}
\label{fig:bw_optsig}
\end{figure}

Then we investigate the relationship between $\sigma^*$ (the optimal bandwidth allocation ratio) and $m$.  In each set of computation, we use \eqref{eq:totalbw}, \eqref{eq:sigma} and the equation array derived in \sref{sec:analysis}, compute the optimal ratio $\sigma^*$ by search in each series of ($\sigma$, $p_{co}$) for each $m$. Also, a larger range of $m$ (1..25) and a smaller step size (one) are used. We perform multiple sets of computation with different $n$ and $\lambda$ corresponding to different network scenarios.

\fref{fig:bw_optsig} presents these results. The first observation is that $\sigma^*$ monotonically increases with $m$. This is consistent with the previous series of ($\A A_1^*$, $\A A_3^*$, ... $\A A_{11}^*$). The second observation is that $\sigma^*$ increases with $n$ (\fref{fig:bw_optsig_diffn}) but decreases with $\lambda$ (\fref{fig:bw_optsig_difflmd}).
This tells us that, to achieve a high availability of cooperation in a sparse network with heavy traffic, the control channel should be allocated much smaller bandwidth than in a dense network with light traffic. ({\bf Finding 7})

\fref{fig:bw_optsig} also shows that $\sigma^*>1$ in most cases. {\red This means that, for a DISH network to achieve larger $p_{co}$, it generally prefers larger bandwidth for the control channel than for {\em each} data channel.} This is contrary to the prior approach of using a smaller frequency band for control (\cite{kya05broadnets,xue06tmc})
or dividing total bandwidth equally among all channels (numerous studies) in non-cooperative multi-channel networks. ({\bf Finding 8})

\ifdefined\thesis
\section{Summary}\label{sec:conc-ana}
DISH represents another dimension of exploiting cooperative diversity, in addition to data relaying, as a control-plane cooperative approach. This chapter %
\else
\section{{\red Discussion and Conclusion}}\label{sec:conc}
{\red Distributed Information SHaring (DISH) represents another dimension of exploiting cooperative diversity, in addition to data relaying, as a distributed flavor of control-plane cooperation.} This paper %
\fi
gives the first theoretical treatment of this notion by addressing the availability of cooperation via a metric $p_{co}$.
Through analysis and investigation of the metric under three contexts, i.e., non-cooperative case, ideal DISH, and real DISH, our study demonstrates that $p_{co}$ is a useful performance indicator for DISH networks and bears significant implications. We also apply our results to solving a practical bandwidth allocation problem.

{\red Our study yields eight findings as listed in \sref{sec:intro-ana}, which serve for different purposes. Findings 1 and 2 back up the feasibility and benefit of DISH. Findings 3 and 4 give hints on improving system performance by adjusting packet size, node density and traffic load. Finding 5 demonstrates the significance of the metric $p_{co}$. Findings 7 and 8 suggest ways of performance improvement from a system design perspective.}

{\red In the case of {\em mobile} ad hoc networks, a node $v$ cannot become a cooperative node with respect to nodes $x$ and $y$ who create an MCC problem, if $v$ fails to decode at least one control message from $x$ and $y$ due to its mobility. However, in most cases when the mobility level is not high, node $v$ will still remain in the intersection region of $x$ and $y$ during the period $x$ and $y$ are sending their control messages, and thus is still able to cooperate. Also, another effect can compensate for the missing of cooperative nodes: a node, although not in the intersection range of $x$ and $y$, first hears $x$'s control message and then moves into $y$'s range and hears $y$'s control message. In this case, it can also identify the MCC problem and become a cooperative node.}

This work attempts to encourage an insightful understanding of DISH, and based on our findings, it demonstrates that $p_{co}$ is a useful metric capable of characterizing the performance of DISH networks, and also bears significant implications. We contend that DISH, as a new cooperative approach, is practical enough to be a part of future cooperative communication networks.

\appendices
\section{Remark on data channel sojourn time}
In our system model specified in \sref{sec:model-ana}, we assume that, after switching to a data channel, a node will stay on that channel for $T_d$, the duration of a successful data channel handshake.
This is also valid for a failed data channel handshake. To elaborate, let $J_t$ and $J_r$ be the period of staying on a data channel for a transmitter and a receiver, respectively, and let $T_{pkt}$, $T_{pkt}^{tmo}$ and $T_{pkt}^{det}$ be the transmission time of a $pkt$ (\texttt{DATA} or \texttt{ACK}), the timeout interval for receiving a $pkt$, and the interval for detecting $pkt$ collision, respectively. Hence clearly, $T_d=T_{data}+T_{ack}$. For a failed data channel handshake, there are three possible cases: (i) \texttt{DATA} is collided, in which case $J_t=T_{data}+T_{ack}^{tmo}$ and $J_r=T_{data}^{det}$, (ii) \texttt{ACK} is collided, in which case $J_t=T_{data}+T_{ack}^{det}$ and $J_r=T_d$, or (iii) only the receiver switches to the data channel (\texttt{McCTS} fails to reach the transmitter), in which case $J_r=T_{data}^{tmo}$. In all above cases, $J_t\approx J_r\approx T_d=T_{data}+T_{ack}$, because $T_{data}^{tmo}\approx T_{data}\gg T_{ack}^{tmo}$, and $T_{pkt}\approx T_{pkt}^{det}$ (collision detection is typically by checking CRC at the footer of a packet).

\section{}
\subsection{{\red Proof of Proposition~\ref{prop:notintfoh}}}
\begin{lem} \label{lem:switch}
If node $u$ is on a data channel at $t_1$, then the probability that $u$ does not introduce interference to the control channel during $[t_1,t_2]$, where $t_2-t_1=\Delta t<T_d$, is given by
\begin{align*}
 \Pr[\A I_u(t_1,t_2)| \overline{\A C_u(t_1)}]
=1-\frac{\Delta t}{T_d} + \frac{1- e^{-\lambda_c \Delta t}}{\lambda_c T_d}.
\end{align*}
\end{lem}

\begin{IEEEproof}
By the total probability theorem,
\ifdefined\thesis
\begin{equation*}
l.h.s.=\Pr[\Omega_u(t_1,t_2)]\times\Pr[\A I_u(t_1,t_2) |\Omega_u(t_1,t_2)]+\Pr[\overline{\Omega_u(t_1,t_2)}]\times 1.
\end{equation*}
\else
\begin{align*}
l.h.s.=&\Pr[\Omega_u(t_1,t_2)]\times\Pr[\A I_u(t_1,t_2) |\Omega_u(t_1,t_2)]\\ &+\Pr[\overline{\Omega_u(t_1,t_2)}]\times 1.
\end{align*}
\fi

Let $t_{sw}$ be the time when node $u$ switches to the control channel (see \fref{fig:switch}). It is uniformly distributed in $[t_1, t_1+T_d]$ because the time when $u$ started its data channel handshake is unknown, and hence
\begin{align}\label{eq:switch}
 \Pr[\Omega_u(t_1,t_2)] = \frac{\Delta t}{T_d}.
\end{align}

\begin{figure}[tb]
\centering
  \includegraphics[trim=0 0 0 3mm,clip,width=0.75\linewidth]{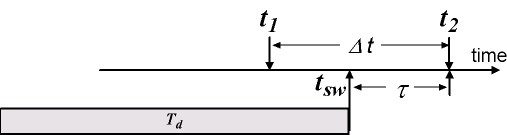}
  \caption{A node switches to the control channel after data channel handshaking.}
  \label{fig:switch}
\vspace{-3mm}\end{figure}

Since control channel traffic is Poisson with rate $\lambda_c$,
\begin{align*}
\Pr[\A I_u(t_1, t_2) | \Omega_u(t_1,t_2)]
  =\Pr[\A S_u(t_{sw}, t_2) | \Omega_u(t_1,t_2)]
  =\bb E [e^{-\lambda_c \tau}]
\end{align*}
where $\tau=t_2 - t_{sw}$ is uniformly distributed in $[0,\Delta t]$ by the same argument leading to \eqref{eq:switch}. Hence
\begin{align*}
\mathbb{E} [e^{-\lambda_c \tau}]
    = \int_0^{\Delta t} e^{-\lambda_c \tau_0} \frac{1}{\Delta t} d\tau_0
    = \frac{1- e^{- \lambda_c \Delta t}}{\lambda_c \Delta t},
\end{align*}
and then by substitution the lemma is proven.
\end{IEEEproof}

The proof of \pref{prop:notintfoh} is as below.
\begin{IEEEproof}
In the case of $u \in \A N_{vi}$, no matter $u$ is on the control channel at $s_i$, or is on a data channel at $s_i$ but switches to the control channel before $s_i+b$, it will sense a busy control channel (due to CSMA) and thus keep silent.

\begin{figure}[tb]
\centering\includegraphics[width=0.88\linewidth]{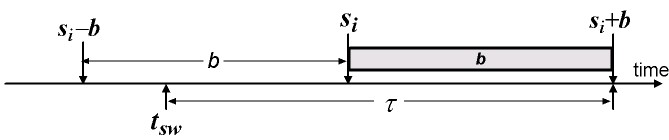}
\caption{The vulnerable period of $v$ is $[s_i-b,s_i+b]$, in which node $u\in \A N_{v\backslash i}$ should not start transmission on the control channel.}
\label{fig:vp}
\end{figure}

In the case of $u \in \A N_{v\backslash i}$, see \fref{fig:vp}. Note that the vulnerable period of $v$ is $[s_i-b, s_i+b]$ instead of $[s_i, s_i+b]$, because a transmission started within $[s_i-b,s_i]$ will end within $[s_i, s_i+b]$. Therefore, by the total probability theorem,
\begin{align*}
p_{ni\text{-}oh}
 =& \Pr[\A C_u(s_i-b)] \cdot \Pr[\A I_u(s_i-b, s_i+b)|\A C_u(s_i-b)]\\
  &+\Pr[\overline{\A C_u(s_i-b)}] \cdot \Pr[\A I_u(s_i-b, s_i+b) |\overline{\A C_u(s_i-b)}]\\
 =&\ p_{ctrl} \cdot e^{-2 \lambda_c b} +(1-p_{ctrl})
  \cdot (1-\frac{2b}{T_d} + \frac{1- e^{-2 \lambda_c b}}{\lambda_c T_d}),
\end{align*}
where $\Pr[\A I_u(s_i-b, s_i+b) |\overline{\A C_u(s_i-b)}]$ is solved by \lref{lem:switch}.
\end{IEEEproof}

\subsection{Derivation of \eqref{eq:poh-vi}}
\begin{IEEEproof}
Based on the proof for the case $u \in \A N_{vi}$ in \pref{prop:notintfoh}, it is easy to show that $\A S_v(s_i, s_i+b)\Leftrightarrow\A C_v(s_i)$. Hence
\[ \Pr[\A S_v(s_i, s_i+b)] = \Pr[\A C_v(s_i)] = p_{ctrl}. \]
Treating events $\A S_v(s_i, s_i+b)$ (node $v$ is silent on the control channel) and $\A I_u(s_i, s_i+b)$ (node $u$ does not interfere the control channel) being independent of each other, as an approximation,
we have
\[ \Pr[\A O (v\leftarrow i)] \approx p_{ctrl} \prod_{u \in \A N_{v\backslash i}} p_{ni\text{-}oh}
        = p_{ctrl} \; p_{ni\text{-}oh}^{K_{v\backslash i}}. \]
\end{IEEEproof}

\subsection{{\red Proof of Proposition~\ref{prop:notintfcts}}}
\begin{IEEEproof}
The case of $u \in \A N_{ij}$ follows the same line as the proof for \pref{prop:notintfoh}.   In the case of $u \in \A N_{i\backslash j}$, the only difference from \pref{prop:notintfoh} is that now we are implicitly given the fact that $i$ was transmitting \texttt{McRTS} during $[s_j-b, s_j]$. This excludes $i$'s any neighbor $u$ interfering in $[s_j-b, s_j]$. Therefore $i$'s vulnerable period is $[s_j, s_j+b]$ instead of $[s_j-b, s_j+b]$ as compared to \pref{prop:notintfoh}. So
\begin{align*}
p_{ni\text{-}cts} =& \Pr[\A C_u(s_j-b)] \cdot \Pr[\A I_u(s_j, s_j+b)|\A C_u(s_j-b)]\\
 &+\Pr[\overline{\A C_u(s_j-b)}]\cdot \Pr[\A I_u(s_j, s_j+b)|\overline{\A C_u(s_j-b)}].
\end{align*}
Note that we condition on $\A C_u(s_j-b)$ instead of $\A C_u(s_j)$, because $s_j$ is not an {\em arbitrary} time due to $i$'s \texttt{McRTS} transmission during $[s_j-b, s_j]$, which leads to $\Pr[\A C_u(s_j)]\neq p_{ctrl}$.

First, $\Pr[\A I_u(s_j, s_j+b)|\A C_u(s_j-b)]=1$. This is because, as $\A C_u(s_j-b)\Leftrightarrow \A S_u(s_j-b, s_j)$ which is easy to show, $u$ will successfully overhear $i$'s \texttt{McRTS}, and hence will keep silent in the next period of $b$ to avoid interfering with $i$ receiving \texttt{McCTS}.

Next consider $\Pr[\A I_u(s_j, s_j+b)|\overline{\A C_u(s_j-b)}]$ where $u$ is on a data channel at $s_j-b$. If $u$ switches to the control channel (i) before $s_j$, it will be suppressed by $i$'s \texttt{McRTS} transmission until $s_j$, and thus the vulnerable period of $i$ receiving \texttt{McCTS} is $[s_j, s_j+b]$, (ii) within $[s_j, s_j+b]$, this has been solved {\red by \lref{lem:switch}}, or (iii) after $s_j+b$, the probability to solve is obviously 1. Therefore,
\begin{align*}
\Pr[&\A I_u(s_j, s_j+b)|\overline{\A C_u(s_j-b)}]
= \Pr[\Omega_u(s_j-b, s_j)]\; e^{-\lambda_c b}\\
    &+ \Pr[\Omega_u(s_j, s_j+b)]\,(1-\frac{b}{T_d} + \frac{1- e^{-\lambda_c b}}{\lambda_c T_d})\\
    &+ \{1-\Pr[\Omega_u(s_j-b, s_j)]-\Pr[\Omega_u(s_j, s_j+b)]\}\times 1.
\end{align*}
According to \eqref{eq:switch}, $\Pr[\Omega_u(s_j-b, s_j)]=\Pr[\Omega_u(s_j, s_j+b)]=b/T_d$. Then by substitution the proposition is proven.
\end{IEEEproof}

\subsection{{\red Proof of \lref{lem:avgarea}}}
\begin{IEEEproof}
Let $A_s(\gamma)$ be the intersection area of two circles with a distance of $\gamma$ between their centers, and $\gamma<R$, where $R$ is the circles' radius. It can be derived from \cite{mathwd-cirseg} that
\begin{align*}
    A_s(\gamma) = 2 R^2 \arccos\frac{\gamma}{2R} -
        \gamma\sqrt{R^2-\frac{\gamma^2}{4}}.
\end{align*}
Let $A_c(\gamma)$ be the complementary area of $A_s(\gamma)$, i.e., $A_c(\gamma) = \pi R^2 - A_s(\gamma)$,
and let $A_{ij}$ and $A_{v\backslash i}$ be the areas where $\A N_{ij}$ and $\A N_{v\backslash i}$ are located, respectively.

(a) See \fref{fig:distpdf-all}. Letting $\gamma=||vi||$ where $v\in \A N_i$, and $f(r)$ be its probability density function (pdf), we have $f(r) dr = 2 \pi r dr/(\pi R^2)$, which gives $f(r) = 2 r/R^2$. Thus
\begin{align*}
    \mathbb{E}[A_{v\backslash i}|v\in \A N_i]
    = \int_0^R A_c(r) f(r) d r 
    \approx 1.30 R^2,
\end{align*}
and hence $\mathbb{E}[K_{v\backslash i}|v\in \A N_i] \approx n\cdot 1.30 R^2/R^2 = 1.30 n$.

\begin{figure}[tb]
\centering
\subfloat[$v\in \A N_i$.]
    {\includegraphics[width=0.33\linewidth]{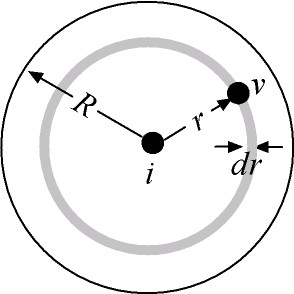}
    \label{fig:distpdf-all}} \hfil
\subfloat[$v\in\A N_{i\backslash j}$.]
    {\includegraphics[width=0.55\linewidth]{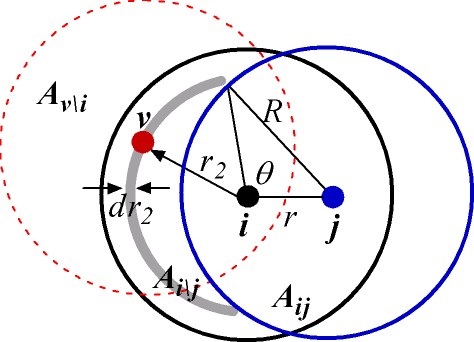}
    \label{fig:distpdf-moon}}
\caption{Deriving the pdf of distance $||vi||$.}
\label{fig:distpdf}
\vspace{-5mm}\end{figure}

(b) Let $\gamma_1=||vi||$ where $v\in \A N_{ij}$, and $f_1(r_1)$ be its pdf.
To solve $f_1(r_1)$, we consider $v\in \A N_{i \backslash j}$ instead (see \fref{fig:distpdf-moon}):
\begin{align}\label{eq:condexp}
&\because\ \mathbb{E}[A_{v\backslash i}|v\in \A N_i]
  = p_1 \cdot \mathbb{E}[A_{v\backslash i}|v\in \A N_{ij}]\notag\\
&\hspace{3.1cm}+(1-p_1) \cdot \mathbb{E}[A_{v\backslash i} | v\in \A N_{i \backslash j}]\notag\\
&\hspace{5mm}\text{where } p_1\triangleq \Pr[v\in \A N_{ij} | v\in\A N_i] = \frac{A_s(r)}{\pi R^2},\notag\\
&\therefore\ \mathbb{E}[A_{v\backslash i} | v\in \A N_{ij}]
    = p_1^{-1}\mathbb{E}[A_{v\backslash i}|v\in \A N_i]\notag\\
&\hspace{3.1cm}- (p_1^{-1}-1)\cdot \mathbb{E}[A_{v\backslash i} | v\in \A N_{i \backslash j}].
\end{align}
To determine $\mathbb{E}[A_{v\backslash i} | v\in \A N_{i\backslash j}]$, let $\gamma_2=||vi||$ where $v\in \A N_{i\backslash j}$, and $f_2(r_2)$ be its pdf. It is determined by
\begin{align*}
f_2(r_2) dr_2 = \frac{2(\pi - \theta)r_2 dr_2}{A_{i\backslash j}} \text{ and }
\cos\theta = \frac{r_2^2 + r^2 - R^2}{2r_2 r}.
\end{align*}
Therefore
\begin{align*}
 &\mathbb{E}[A_{v\backslash i} | v\in \A N_{i\backslash j}]
   = \int_{R-r}^R A_c(r_2) f_2(r_2) dr_2\\
=&\int_{R-r}^R \frac{2 r_2 A_c(r_2)}{A_c(r)}
    (\pi - \arccos \frac{r_2^2 + r^2 - R^2}{2r_2 r} ) dr_2.
\end{align*}
Substituting this and $\mathbb{E}[A_{v\backslash i}|v\in \A N_i]\approx 1.30 R^2$ (by case (a)) into \eqref{eq:condexp} solves $\bb{E}[A_{v\backslash i} | v\in \A N_{ij}]$, which we denote by $M(r)$.
Then we have
\[ \mathbb{E}[K_{v\backslash i}|v\in \A N_{ij}] = \frac{n}{R^2} \int_0^R M(r)f(r)dr \approx 1.19 n. \]

(c) Proven by noticing that $A_{ij}$ is complementary to the area corresponding to case (a).
\end{IEEEproof}

\subsection{Derivation of \eqref{eq:poh_psucc}}
\begin{IEEEproof}
Taking the expectation of $\Pr[\A O (v\leftarrow i)]$ (given by \eqref{eq:poh-vi}) over all neighboring $(v,i)$ pairs using \lref{lem:epk} and \lref{lem:avgarea}-(a):
\begin{align*}
p_{oh} \approx p_{ctrl}\; \bb{E}[p_{ni\text{-}oh}^{K_{v\backslash i}}]
    \approx p_{ctrl}\; \exp[-1.30 n (1-p_{ni\text{-}oh})].
\end{align*}

To solve for $p_{succ}$, notice that for a control channel handshake to be successful, (i) the \texttt{McRTS} must be successfully received by the receiver, with probability $p_{oh}$, and (ii) the \texttt{McCTS} must be successfully received by the transmitter, with probability $\bb E[p_{ni\text{-}cts}^{K_{i\backslash j}}]$ based on \pref{prop:notintfcts} (assuming that $p_{ni\text{-}cts}$ holds for nodes in $\A N_{i\backslash j}$ independently, as an approximation). Therefore,
\begin{align*}
p_{succ} \approx p_{oh} \mathbb{E}[p_{ni\text{-}cts}^{K_{i\backslash j}}]
    \approx p_{oh} \exp[-1.30n(1-p_{ni\text{-}cts})].
\end{align*}
\end{IEEEproof}

\subsection{Derivation of \eqref{eq:pctrlstar}}
\begin{IEEEproof}
Recall that node $v$ must stay continuously on the control channel during $[s_x,s_y]$. Let $\tau_c=s_y-s_x$ and suppose $v$ switches to a data channel at $s_x+\tau_w$, then we need $\tau_w>\tau_c$. Hence $p_{ctrl}^{\star} = \Pr(\tau_w>\tau_c)$, where $\tau_c\in [0,T_d]$.

\begin{figure}[tb]
\centering\includegraphics[width=0.98\linewidth]{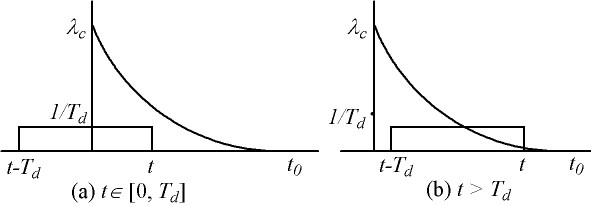}
\caption{The convolution of $\frac{1}{T_d}$ ($t\in [0,T_d]$) and $\lambda_c e^{-\lambda_c t}$ ($t>0$).}
\label{fig:convolute}
\vspace{-5mm}\end{figure}

Denote by $f_{\tau_c}(t)$ the pdf of an unbounded $\tau_c$ ($s_y\in (s_x,\infty)$). The fact that a MCC problem is created by $x$ and $y$ (at $s_y$) implies that $y$ missed $x$'s control message (at $s_x$). This is due to one of the following: (i) $y$ is on the control channel at $s_x$ but interfered, in which case $f_{\tau_c}(t)$ is $\lambda_c e^{-\lambda_c t}$ (ignoring the short interference period which is in the magnitude of $b$, while $\tau_c$ is in the magnitude of $T_d$), (ii) $y$ is on a data channel at $s_x$, in which case $y$ must switch to the control channel before $s_y$. Again see \fref{fig:switch}, where $t_1$ and $t_2$ are now $s_x$ and $s_y$, respectively. Let $\tau_1=t_{sw}-s_x$ and $\tau_2=s_y-t_{sw}$, then $\tau_c=\tau_1+\tau_2$. Note that $\tau_1$ is uniformly distributed in $[0,T_d]$, $\tau_2$ is exponentially distributed with the mean of $1/\lambda_c$, and $\tau_1$ and $\tau_2$ can be regarded as independent. Therefore, $f_{\tau_c}(t)$ is the convolution of $\frac{1}{T_d}$ ($t\in [0,T_d]$) and $\lambda_c e^{-\lambda_c t}$ ($t>0$), which can be calculated by referring to \fref{fig:convolute}, to be $f_{\tau_c}^d (t) =$
\begin{align*}
\frac{1- e^{-\lambda_c t} }{T_d} [u(t)-u(t-T_d)]
         + \frac{ e^{-\lambda_c t} }{T_d} (e^{\lambda_c T_d} -1) u(t-T_d).
\end{align*}
where $u(\cdot)$ is the unit step function.

A weighted sum of the above cases (i) and (ii) gives
\begin{align*}
  f_{\tau_c}(t) = w\; \lambda_c e^{-\lambda_c t} + (1-w)\; f_{\tau_c}^d (t)
\end{align*}
where $w$ is the weight for case (i). To determine $w$, note that the probability of case (ii) is $1-p_{ctrl}$, and the probability of case (i) is $p_{ctrl} (1-p_{ni\text{-}oh}^{K_{y\backslash x}})$ (using \eqref{eq:poh-vi}) whose mean is $ p_{ctrl} (1-\exp[-1.30 n(1-p_{ni\text{-}oh})])$. Therefore
\begin{align*}
w =\frac{p_{ctrl} [1-e^{-1.30n(1-p_{ni\text{-}oh})}]}
{p_{ctrl}[1-e^{-1.30n(1-p_{ni\text{-}oh})}] + (1-p_{ctrl})}
  = \frac{p_{ctrl}-p_{oh}}{1-p_{oh}}.
\end{align*}

Finally we compute $p_{ctrl}^{\star} = \Pr(\tau_w>\tau_c)$ using $f_{\tau_c}(t)$. Recall that $f_{\tau_c}(t)$ is the pdf of an unbounded $\tau_c$ but $\tau_c$ is in fact bounded within $[0,T_d]$, therefore its actual pdf is\\
$f_{\tau_c}(t) / \int_0^{T_d} f_{\tau_c}(t) dt$. Assuming that $\tau_w$ is exponentially distributed with mean $1/\lambda_w$, we have
\begin{align*}
p_{ctrl}^{\star} = \bb E_{\tau_c\in[0,T_d]}\;\Pr(\tau_w>\tau_c)
    = \int_0^{T_d} e^{-\lambda_w t} \frac{f_{\tau_c}(t)}{\int_0^{T_d} f_{\tau_c}(t) dt} dt
\end{align*}
which reduces to \eqref{eq:pctrlstar}.  For $\lambda_w$, noticing that it is the average rate of a node on the control channel switching to data channels, which happens when a node successfully initiates a control channel handshake via \texttt{McRTS} or sends a \texttt{McCTS}, we have $\lambda_w=\lambda_{rts} p_{succ}+\lambda_{cts}$.
\end{IEEEproof}

\bibliographystyle{IEEEtran}
\bibliography{IEEEabrv,../../common/references}

\begin{thebibliography}{10}
\providecommand{\url}[1]{#1}
\csname url@samestyle\endcsname
\providecommand{\newblock}{\relax}
\providecommand{\bibinfo}[2]{#2}
\providecommand{\BIBentrySTDinterwordspacing}{\spaceskip=0pt\relax}
\providecommand{\BIBentryALTinterwordstretchfactor}{4}
\providecommand{\BIBentryALTinterwordspacing}{\spaceskip=\fontdimen2\font plus
\BIBentryALTinterwordstretchfactor\fontdimen3\font minus
  \fontdimen4\font\relax}
\providecommand{\BIBforeignlanguage}[2]{{%
\expandafter\ifx\csname l@#1\endcsname\relax
\typeout{** WARNING: IEEEtran.bst: No hyphenation pattern has been}%
\typeout{** loaded for the language `#1'. Using the pattern for}%
\typeout{** the default language instead.}%
\else
\language=\csname l@#1\endcsname
\fi
#2}}
\providecommand{\BIBdecl}{\relax}
\BIBdecl

\bibitem{van3term}
E.~C. van~der Meulen, ``Three-terminal communication channels,'' \emph{Advances
  in Applied Probability}, vol.~3, no.~1, pp. 120--154, 1971.

\bibitem{cover79}
T.~M. Cover and A.~E. Gamal, ``Capacity theorems for the relay channel,''
  \emph{IEEE Trans. Infomation Theory}, vol.~25, pp. 572--84, Sept. 1979.

\bibitem{coopcap06}
A.~Host-Madsen, ``Capacity bounds for cooperative diversity,'' \emph{IEEE
  Trans. on Information Theory}, vol.~52, no.~4, pp. 1522--1544, Apr. 2006.

\bibitem{laneman03}
J.~N. Laneman and G.~W. Wornell, ``Distributed space-time-coded protocols for
  exploiting cooperative diversity in wireless networks,'' \emph{IEEE
  Transactions on Information Theory}, vol.~49, no.~10, October 2003.

\bibitem{coop03toc1}
A.~Sendonaris, E.~Erkip, and B.~Aazhang, ``User cooperation diversity--part
  {I}: System description,'' \emph{IEEE Trans. Commun.}, vol.~51, no.~11,
  December 2003.

\bibitem{coop03toc2}
------, ``User cooperation diversity--part {II}: Implementation aspects and
  performance analysis,'' \emph{IEEE Trans. Commun.}, vol.~51, no.~11, Dec.
  2003.

\bibitem{rdcf05infocom}
H.~Zhu and G.~Cao, ``{rDCF}: A relay-enabled medium access control protocol for
  wireless ad hoc networks,'' in \emph{IEEE Infocom}, 2005.

\bibitem{cmac05globecom}
A.~Azgin, Y.~Altunbasak, and G.~AlRegib, ``Cooperative {MAC} and routing
  protocols for wireless ad hoc networks,'' in \emph{IEEE GLOBECOM}, 2005.

\bibitem{cdmac07icc}
S.~Moh, C.~Yu, S.-M. Park, and H.-N. Kim, ``{CD-MAC}: Cooperative diversity
  {MAC} for robust communication in wireless ad hoc networks,'' in \emph{IEEE
  ICC}, June 2007, pp. 3636--3641.

\bibitem{comac07jsac}
P.~Liu, Z.~Tao, S.~Narayanan, T.~Korakis, and S.~S. Panwar, ``{CoopMAC}: A
  cooperative {MAC} for wireless {LANs},'' \emph{IEEE Journal On Selected Areas
  In Communications}, vol.~25, no.~2, pp. 340--354, 2007.

\bibitem{tie06cam}
T.~Luo, M.~Motani, and V.~Srinivasan, ``{CAM-MAC}: A cooperative asynchronous
  multi-channel {MAC} protocol for ad hoc networks,'' in \emph{IEEE Broadnets},
  San Jose, CA, USA, October 2006.

\bibitem{tie09tmc}
------, ``Cooperative asynchronous multichannel {MAC}: Design, analysis, and
  implementation,'' \emph{IEEE Transactions on Mobile Computing}, vol.~8,
  no.~3, pp. 338--52, March 2009.

\bibitem{dca00}
S.-L. Wu, C.-Y. Lin, Y.-C. Tseng, and J.-P. Sheu, ``A new multi-channel {MAC}
  protocol with on-demand channel assignment for multi-hop mobile ad hoc
  networks,'' in \emph{I-SPAN}, 2000.

\bibitem{nas99}
A.~Nasipuri, J.~Zhuang, and S.~R. Das, ``A multichannel {CSMA} {MAC} protocol
  for multihop wireless networks,'' in \emph{WCNC}, 1999.

\bibitem{jain01}
N.~Jain, S.~R. Das, and A.~Nasipuri, ``A multichannel {CSMA} {MAC} protocol
  with receiver-based channel selection for multihop wireless networks,'' in
  \emph{IEEE ICCCN}, 2001.

\bibitem{mup04}
A.~Adya, P.~Bahl, J.~Padhye, and A.~Wolman, ``A multi-radio unification
  protocol for {IEEE} 802.11 wireless networks,'' in \emph{IEEE Broadnets},
  2004.

\bibitem{mah06}
R.~Maheshwari, H.~Gupta, and S.~R. Das, ``Multichannel {MAC} protocols for
  wireless networks,'' in \emph{IEEE SECON}, 2006.

\bibitem{chen03}
J.~Chen, S.~Sheu, and C.~Yang, ``A new multichannel access protocol for {IEEE}
  802.11 ad hoc wireless {LANs},'' in \emph{PIMRC}, 2003.

\bibitem{mmac04}
J.~So and N.~Vaidya, ``Multi-channel {MAC} for ad hoc networks: Handling
  multi-channel hidden terminals using a single transceiver,'' in \emph{ACM
  MobiHoc}, 2004.

\bibitem{tmmac07}
J.~Zhang, G.~Zhou, C.~Huang, S.~H. Son, and J.~A. Stankovic, ``{TMMAC}: an
  energy efficient multi-channel {MAC} protocol for ad hoc networks,'' in
  \emph{IEEE ICC}, 2007.

\bibitem{chma00}
A.~Tzamaloukas and J.~Garcia-Luna-Aceves, ``Channel-hopping multiple access,''
  in \emph{IEEE ICC}, 2000.

\bibitem{chat00}
------, ``Channel-hopping multiple access with packet trains for ad hoc
  networks,'' in \emph{IEEE Device Multimedia Communications}, 2000.

\bibitem{ssch04}
P.~Bahl, R.~Chandra, and J.~Dunagan, ``{SSCH}: Slotted seeded channel hopping
  for capacity improvement in {IEEE} 802.11 ad-hoc wireless networks,'' in
  \emph{ACM MobiCom}, 2004.

\bibitem{tie08mobihoc}
T.~Luo, M.~Motani, and V.~Srinivasan, ``Analyzing {DISH} for multi-channel
  {MAC} protocols in wireless networks,'' in \emph{ACM MobiHoc}, Hong Kong,
  China, 2008.

\bibitem{han05}
Y.~S. Han, J.~Deng, and Z.~J. Haas, ``Analyzing multi-channel medium access
  control schemes with {ALOHA} reservation,'' \emph{IEEE Trans. on Wireless
  Comm.}, 2005.

\bibitem{tie07mobicom}
T.~Luo, M.~Motani, and V.~Srinivasan, ``Altruistic cooperation for
  energy-efficient multi-channel {MAC} protocols,'' in \emph{ACM MobiCom},
  Montreal, QC, Canada, 2007.

\bibitem{xue04conn}
F.~Xue and P.~R. Kumar, ``The number of neighbors needed for connectivity of
  wireless networks,'' \emph{Wireless Networks}, vol.~10, no.~2, pp. 169--181,
  2004.

\bibitem{fama95}
C.~L. Fullmer and J.~J. Garcia-Luna-Aceves, ``Floor acquisition multiple access
  {(FAMA)} for packet-radio networks,'' in \emph{SIGCOMM}, New York, NY, USA,
  1995.

\bibitem{fama99}
J.~J. Garcia-Luna-Aceves and C.~L. Fullmer, ``Floor acquisition multiple access
  {(FAMA)} in single-channel wireless networks,'' \emph{Mobile Networks and
  Applications}, vol.~4, no.~3, pp. 157--174, 1999.

\bibitem{kya05broadnets}
P.~Kyasanur, J.~Padhye, and P.~Bahl, ``On the efficacy of separating control
  and data into different frequency bands,'' in \emph{IEEE Broadnets}, October
  2005.

\bibitem{xue06tmc}
X.~Yang, N.~H. Vaidya, and P.~Ravichandran, ``Split-channel pipelined packet
  scheduling for wireless networks,'' \emph{IEEE Transactions on Mobile
  Computing}, vol.~5, no.~3, pp. 240--257, 2006.

\bibitem{mathwd-cirseg}
E.~W. Weisstein, ``Circular segment---from mathworld,''
  http://mathworld.wolfram.com/CircularSegment.html.

\end{thebibliography}

\end{document}